\documentclass[9pt,twocolumn,twoside]{osajnl}

\journal{ao} 

\setboolean{shortarticle}{false} 

\usepackage{subcaption}
\usepackage{varwidth}
\usepackage[normalem]{ulem}
\usepackage{multirow}
\usepackage{threeparttable}

\title{A 1.6:1 Bandwidth Two-Layer Antireflection Structure for Silicon Matched to the 190--310~GHz Atmospheric Window}

\author[1,*]{Fabien~Defrance}
\author[2]{Cecile~Jung-Kubiak}
\author[1]{Jack Sayers}
\author[3]{Jake~Connors}
\author[4]{Clare deYoung}
\author[1]{Matthew~I.~Hollister}
\author[1]{Hiroshige~Yoshida}
\author[2]{Goutam~Chattopadhyay}
\author[1]{Sunil~R.~Golwala}
\author[4]{Simon~J.~E.~Radford}

\affil[1]{Division of Physics, Mathematics, and Astronomy, California Institute of Technology, Pasadena, CA 91125}
\affil[2]{Jet Propulsion Laboratory, California Institute of Technology, Pasadena, CA 91109}
\affil[3]{Harvard-Smithsonian Center for Astrophysics, Cambridge, MA 02138}
\affil[4]{Smithsonian Astrophysical Observatory, Submillimeter Array, Hilo, HI 96720}

\affil[*]{Corresponding author: fdefranc@caltech.edu}

\ociscodes{(050.6624) Subwavelength structures; (220.4000)   Microstructure fabrication; (220.4610) Optical fabrication; (310.6628) Subwavelength structures, nanostructures (350.1260) Astronomical optics}

\begin{abstract}

Although high-resistivity, low-loss silicon is an excellent material for THz transmission optics, its high refractive index necessitates antireflection treatment. We fabricated a wide-bandwidth, two-layer antireflection treatment by cutting subwavelength structures into the silicon surface using multi-depth deep reactive ion etching (DRIE). A wafer with this treatment on both sides has $<$$-$20~dB ($<$1\%) reflectance over 187--317~GHz at 15\textdegree\ angle of incidence in TE polarization. We also demonstrated that bonding wafers introduces no reflection features above the $-$20~dB level (also in TE at 15\textdegree), reproducing previous work. Together these developments immediately enable construction of wide-bandwidth silicon vacuum windows and represent two important steps toward gradient-index silicon optics with integral broadband antireflection treatment.

\end{abstract}

\setboolean{displaycopyright}{true}

\begin{document}

\maketitle

\section{Introduction}

Optical systems at frequencies of tens of GHz to a few THz (a range termed ``THz'' here for brevity) benefit greatly from the use of silicon: its high index of refraction minimizes the thickness and curvature of optical elements; its refractive index is achromatic; it is not birefringent; its low loss, even at room temperature, ensures high optical efficiency can be maintained; its high thermal conductivity ensures it can be cooled effectively in cryogenic applications; and its strength permits it to be used for vacuum windows. Specific THz applications include security imaging, remote sensing, and astronomical observations of the cosmic microwave background and cold, dusty sources such as star-forming regions and dust-obscured galaxies. The high refractive index of silicon, however, necessitates antireflection (AR) treatment. The simplest approaches, quarter-wavelength laminated coatings or etched structures, have only narrow bandwidths, roughly $1.2$:1 ($\nu_{\rm max}$:$\nu_{\rm min}$) with less than 1\% reflectance. Furthermore, laminated coatings require care to avoid mechanical failure in applications where the optics are cooled. There is, therefore, a need in the THz regime for a robust, broadband ($>$2:1) AR treatment technique for silicon optics.  

To this end, we are undertaking a multi-phase program to develop broadband, AR-treated silicon optics for THz frequencies.  The  long-term goal is to construct silicon optics by stacking and wafer-bonding individual silicon wafers, each about 1~mm thick, that have each been patterned using deep reactive-ion etching (DRIE).  
In this paper, we demonstrate two key first steps: use of a multi-depth etch process to fabricate a two-layer AR structure in silicon with $<1$\% ($<$$-$20~dB) reflectance over a 1.6:1 bandwidth; and, use of wafer bonding to stack wafers with no measurable degradation in reflectance.   

We quote all reflectances and transmittances in power (intensity, not electric field) in this work.

\subsection{Previous Work}

The prototypical AR treatment has a quarter-wavelength layer of dielectric material with refractive index equal to the square root of that of the substrate. Broader bandwidths can be obtained by using multiple quarter-wavelength layers with properly selected indices.  To maintain the advantages of silicon, any AR treatment must have low loss, must lack birefringence, and, for cryogenic use, must closely match the thermal contraction of silicon. Few materials meet all these requirements and also have the correct indices. Plastic coatings, such as parylene, are often used in narrow-bandwidth applications with modest reflectance requirements~\cite{Gatesman:00}.  Cirlex has also been used in this fashion~\cite{Lau:2006}.  Another approach is an epoxy-based coating in which each layer's dielectric constant is tuned by mixing different types of epoxy or doping with strontium titanate~\cite{Rosen:13}. This technique has been used to achieve wider bandwidths, yielding a reflectance of $< 10\%$ with bandwidths of 2.7:1 and 3.2:1 in two and three-layer coatings, respectively. The epoxy suffers, however, significant absorption loss: 1\% for two layers and 10\% for three.  Furthermore, $<$1\% reflectance is the typical requirement.  Other approaches that have been developed include plasma spray coatings~\cite{Jeong:16} and artificial dielectric metamaterials~\cite{Zhang:09, Moseley:17}, both of which have similar performance for similar bandwidths.

\begin{figure}[t!]
    \centering
    \includegraphics[width=\linewidth]{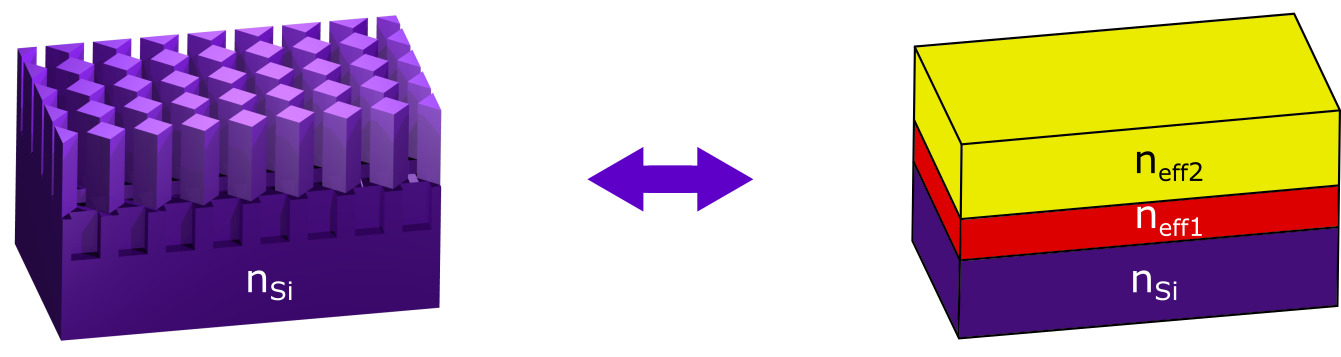}
	\caption{Example of a two-layer AR structure with two different effective refractive indices}
    \label{fig:AR_principle}
\end{figure}

An alternative to conventional dielectrics is to reduce silicon's effective index of refraction by creating sub-wavelength features in its surface (see Fig.~\ref{fig:AR_principle} for an illustration). Varying the microstructure geometry tunes the refractive index.  The loss and thermal contraction requirements are inherently addressed and such microstructures can easily be designed to be non-birefringent at normal incidence.  (No AR approach based on quarter-wavelength structures can be non-birefringent at non-normal incidence.)  

One method to produce such microstructures is to cut them using a dicing saw.  This technique was used to cut crossed grooves in silicon planoconvex lenses for the ACTpol experiment at 150~GHz \cite{Datta:13}, producing a two-layer AR structure with $<$$-$23~dB reflectance over a 1.3:1 bandwidth. The same group has a prototype of a five-layer structure~\cite{CMB-S4_Technology} for which measurements have not been reported. Smooth-sided pyramids cut with a beveled dicing saw have been used to obtain $<5$\% reflectance over a 2.9:1 bandwidth \cite{Young:17}.  In all realizations, the dicing saw approach is limited to producing AR structures that consist of crossed grooves, yielding only posts, and feature sizes are limited by practical saw blade thicknesses. 

Another method is laser machining. In one demonstration, circular holes were bored into a flat alumina sample using a laser, giving $<$10\% reflectance over a 1.3:1 bandwidth~\cite{Nitta:14}. (Alumina is lossier than silicon but has a similar refractive index.)  Other structures, such as sharp cones~\cite{Her:1998}, concentric circular grooves~\cite{Drouet_d'Aubigny:01}, and pyramids~\cite{Matsumura:16} have also been made with laser machining.  It is, however, difficult to control depth with laser machining, and the process can be unacceptably slow for production fabrication.

Extending to THz frequencies and increasing the bandwidth requires finer features than conventional machining can produce. An alternative is to use a photolithographic process to etch features into the silicon surface. DRIE is a mature micromachining technique that can create arbitrary patterns of deep features with aspect ratios up to 30:1.  DRIE has been used in a few demonstrations of flat one-layer structures at THz frequencies~\cite{Gallardo:17,Wheeler:14,Wada:2010,Wagner-Gentner:06,Schuster:05,Yu:17}.  
Multi-layer structures have also been designed and fabricated at THz frequencies using DRIE.  In one case, the design from \cite{Datta:13}, consisting of two layers of posts, was scaled to 850~GHz and fabricated, but test results were not presented~\cite{Gallardo:17}.  In a second case, a structure using three layers of holes was demonstrated with $<$4\% reflectance over a 2.2:1 bandwidth (2.5–5.55~THz)~\cite{Chen:14}.


DRIE, however, is not easily applicable to the curved surfaces of powered optics and seems to have only been demonstrated on such surfaces at much shorter wavelengths~\cite{Kamizuka:14}.
It is thought that slumping techniques may be applicable~\cite{Nakajima:04,Tong:99}, though a demonstration is not forthcoming.

\subsection{A New Approach}
\label{sS:approach}

Combining multi-depth etching with wafer bonding may provide a viable technique for broadband AR treatment for powered silicon optics~\cite{Makitsubo:17}.  A multi-depth DRIE technique has been previously demonstrated~\cite{Jung-Kubiak:16}, providing a means to pattern layers with different refractive indices in a single wafer.  Multiple wafers could then be bonded together to obtain thick, high layer-count structures needed for very wide-bandwidth AR treatments.  To produce powered optics, similar techniques could be used to create a flat-faced gradient-index (GRIN) optic, circumventing the challenge of AR-treating a curved surface.  A cylindrical optic with a parabolic radial index gradient $n(r) = n_0 - r^2/(2 f t_0)$, where $n_0$ is the bulk index, $f$ is the focal length, and $t_0$ is the thickness provides the same focusing as a conventional parabolic lens. The radial index gradient could be achieved by varying the DRIE pattern across each silicon wafer.  Then, as with the AR structure, several wafers could be bonded together to form a focusing optic of the desired thickness. The AR structure would be integrated into the outer layers of the optic, including the variation of the AR treatment with radius.  This AR-textured, GRIN silicon optic would thus present a complete solution to the problem of constructing broadband, powered silicon optical elements.

We consider only structures that exhibit no birefringence at normal incidence.  While even non-birefringent media exhibit polarization-dependent reflectance for non-normal incidence, such effects are minimized if the medium does not exhibit these effects at normal incidence.
Because square grids of circular features or four-fold-symmetric features (e.g., squares, crosses) are non-birefringent, we  chose these as the basis for our design work.  There are arguments that any structure with $N$-fold symmetry with $N > 2$ would be satisfactory~\cite{Mackay:1989}.  Certainly, though, a natural extension of this approach to realize optical elements with useful polarization properties such as quarter- and half-wave plates would be to incorporate such birefringent structures.  Such designs have been implemented using the dicing saw technique~\cite{CMB-S4_Technology}.
  
This paper demonstrates two key aspects of this new approach: multi-depth etching over areas and of structures relevant for modern THz optical systems (100-mm diameter wafers here), and bonding of wafers patterned in this way.  Future work will seek to demonstrate larger bandwidth/layer-count structures and AR-textured GRIN optical elements.

\section{Design}

Our design objective was an optic with two parallel faces, as would be used for a vacuum window, having $<$1\% ($<$$-$20~dB) reflectance across the 190--310~GHz atmospheric window (1.6:1 bandwidth).  Our design process consisted of multiple steps.  First, we applied the well known theory of optical thin films (e.g.,~\cite{BornWolf}) and the equivalent theory of transmission-line impedance transformers (e.g.,~\cite{Pozar}) to design AR structures with multiple quarter-wavelength layers. Second, we used finite-element analysis of one-layer microstructured silicon patterns to determine effective indices of refraction, enabling us to choose appropriate etch patterns for each layer.  Third, we performed a finite-element analysis of the entire multi-layer optic to verify the performance and for comparison with measurements. 
While it would be appropriate to use the finite-element analysis of the full structure to optimize the design parameters, we did not deem that refinement necessary for this demonstration.  Finally, we analyzed the impact of expected fabrication nonidealities. 

We note that any interface between materials having different indices is birefringent for non-normal incidence.  The level of birefringence is generally small for angles less than 30\textdegree~\cite{Schuster:05}.  Since our goal is transmissive optics such as windows and lenses, we deemed this restriction on incidence angle acceptable.  

\begin{figure}[t!]
    \centering
    \includegraphics[width=\linewidth]{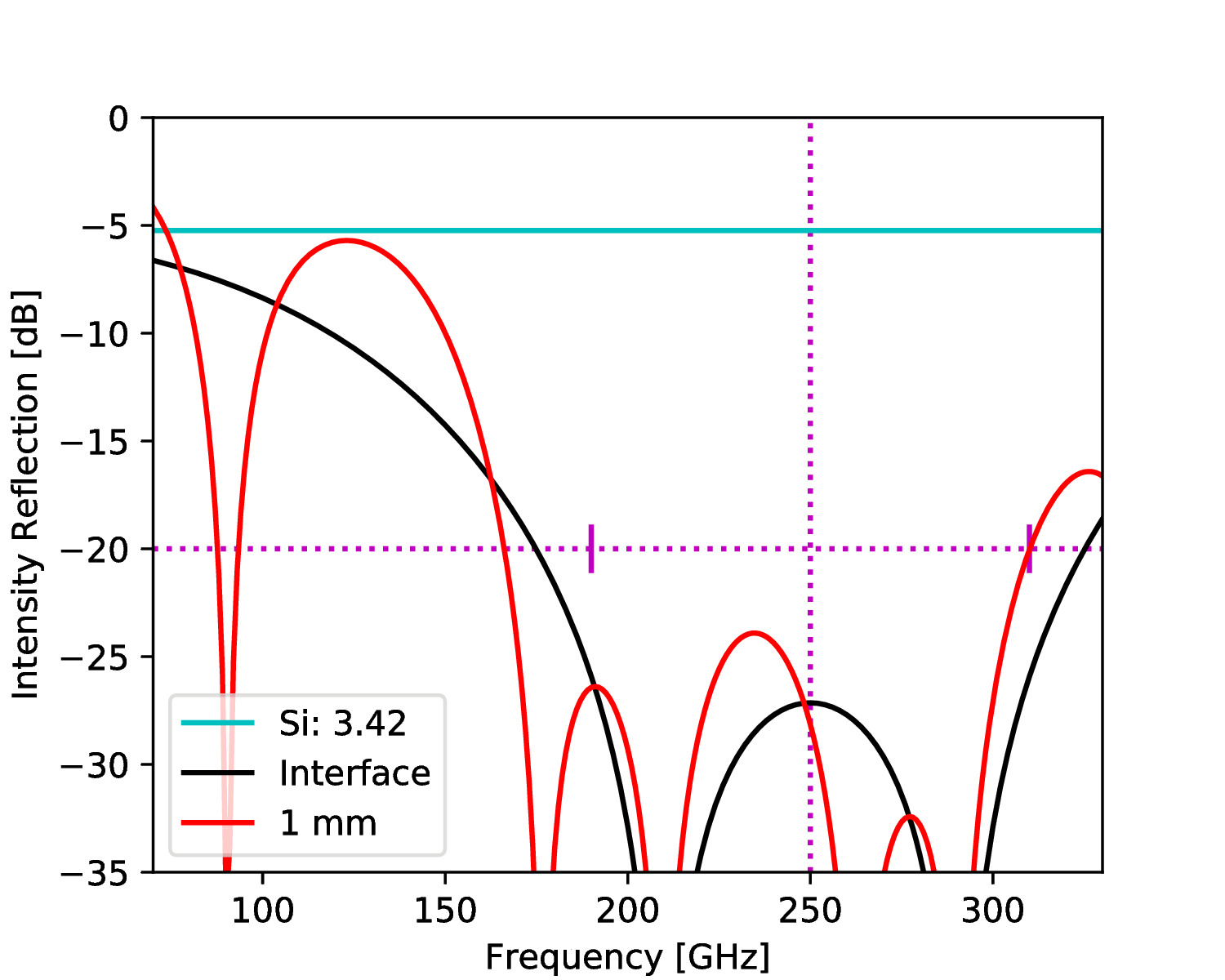}
	\caption{
Two-layer Chebyshev antireflection treatment for silicon. Cyan solid line: A bare silicon surface has a reflectance of 30\% ($- 5.2$~dB). Black solid line: At a single interface, our two-layer design ($n = 1.39$ \& 2.46) provides $< -26$~dB reflectance over the design band, 190--310 GHz. Red solid line: Fabry-P\'erot fringing slightly changes the passband and raises the peak in-band reflectance for a two-sided optic. Even so, it meets the design goal of $< -20$~dB reflectance across the passband. The total wafer thickness of 1 mm includes AR treatment on both sides. }
    \label{fig:AR_design}
\end{figure}

\subsection{Multilayer AR Design}
\label{sS:AR_design}

Silicon's high refractive index, $n_{\rm Si} = 3.42$ at microwave and THz frequencies~\cite{Grischkowsky:90, Dai:04}, causes 30\% reflectance at the interface between vacuum and a bare Si surface. For a single AR layer, the well-known optimal design has an optical thickness of one quarter of the desired wavelength and $n = n_{\rm Si}^{1/2}$, causing reflections from the front and the rear surfaces of the AR layer to destructively interfere at the design frequency and odd harmonics.  A single layer, however, only provides a relatively narrow bandwidth: 1.2:1 at $-$20~dB reflectance and 1.1:1 at $-$26~dB.

Moreover, for a two-sided optic with parallel faces, such as a vacuum window, there will be constructive interference whenever the total optical thickness is an integer number of half wavelengths. To meet the design goal of $<$$-$20~dB reflectance for the entire optic, the reflectance from each single surface must be $<$$-$26~dB to accommodate this 6~dB Fabry-P\'erot fringing. 
For a powered optic (curved or gradient index), we expect the Fabry-P\'erot fringing to be no worse than the above 6~dB, and thus this $-$26~dB criterion is conservative and should be sufficient for powered optics also.

Multi-layer designs provide wider bandwidths. Of the many possibilities, we focused on configurations in which all the layers have quarter-wavelength optical thicknesses. Not only is this approach well studied theoretically, it is well matched to our fabrication technique. Alternate approaches, including pyramids, involve a thicker overall AR structure, more layers, or both for equivalent performance.  

Analogous to filter design, there is a compromise between the bandwidth and the maximum reflectance in the passband. 
There exists a technique using Chebyshev polynomials that provides, for a required bandwidth and maximum in-band reflectance, the number of layers and their refractive indices needed to meet the requirement while providing uniform ripple through the passband~\cite{Pozar, Baumeister:86}. There are many design tools available, including online calculators (e.g., \url{https://www.microwaves101.com}).  Our requirement of a maximum reflectance of $-$26~dB over the 190--310~GHz band results in a 2-layer design with indices and thicknesses given in Table~\ref{tab:AR_design} and with predicted performance shown in Fig.~\ref{fig:AR_design}.

These AR designs implicitly assume the refractive indices of the layers are achromatic through the frequency range of interest. Although this is true for bulk silicon, it is not strictly true for microstructured silicon. Hence the realization of the AR designs must be verified by finite-element analysis.

\subsection{Effective Index of Microstructured Silicon}
\label{sS:design_neff_hfss}

To realize the AR designs, we need to know the effective refractive index, $n_{\rm eff}$, of microstructured silicon. To our knowledge, however, there is not a comprehensive theory in the literature of the effective refractive index of a microstructured dielectric, even in the zero-frequency (static) limit. Although a few models provide guides for specific configurations, they are not applicable in general and are insufficient for design purposes.

We therefore characterized microstructure geometries by using a commercial electromagnetic finite element solver, ANSYS High Frequency Structure Simulator (HFSS), to calculate the complex reflection and transmission spectra ($S$ parameters). We then fit the spectra with dielectric slab models parameterized by an effective refractive index, $n_{\rm eff}$, and an effective thickness, $t_{\rm eff}$.  For a true ``effective index'' approach, it should not be necessary to specify both parameters.  We follow~\cite{Datta:13} in using these two parameters, presumably as a first-order correction to deviations from a pure effective index model.  We explored both achromatic models with constant parameters and models with a linear frequency dependence. 

We modeled unit cells with microstructure features placed between semi-infinite vacuum and semi-infinite silicon.  We placed the input and output ports at the interfaces to vacuum and bulk silicon. We used periodic boundary conditions at the cell walls to emulate an infinite grid. To avoid diffraction effects, the grid spacing, $\Lambda$, must be considerably smaller than the vacuum wavelength, $\lambda / \Lambda > (n_{\rm Si} + \cos \theta) \approx 4$, for incident angles $\theta \le 35^\circ$~\cite{Morris:1993ir}. This grid spacing is much smaller than the 100-mm wafer diameter, so the infinite array approximation should be valid.

Although we explored a variety of microstructure geometries, including linear grooves, hexagonal grids, and circular features, we restrict our attention here to square features in square grids because we found the other geometries provided no additional design flexibility (aside from undesired birefringence at normal incidence).
We modeled straight-walled features, both posts and holes, as are produced by DRIE. We used feature depths close to one quarter of the wavelength at the target spectral band's center frequency, 250 GHz. We parameterized the patterns by the fill factor, $f_{\rm Si}$, which is the ratio of the silicon area to the unit cell area. We calculated the spectra over the range 50--500 GHz and chose a grid spacing $\Lambda = 125$ $\mu$m to obtain a wavelength-to-grid ratio $48 \ge \lambda / \Lambda \ge 4.8$. 

\begin{figure}[t!]
    \centering
    \includegraphics[width=\linewidth]{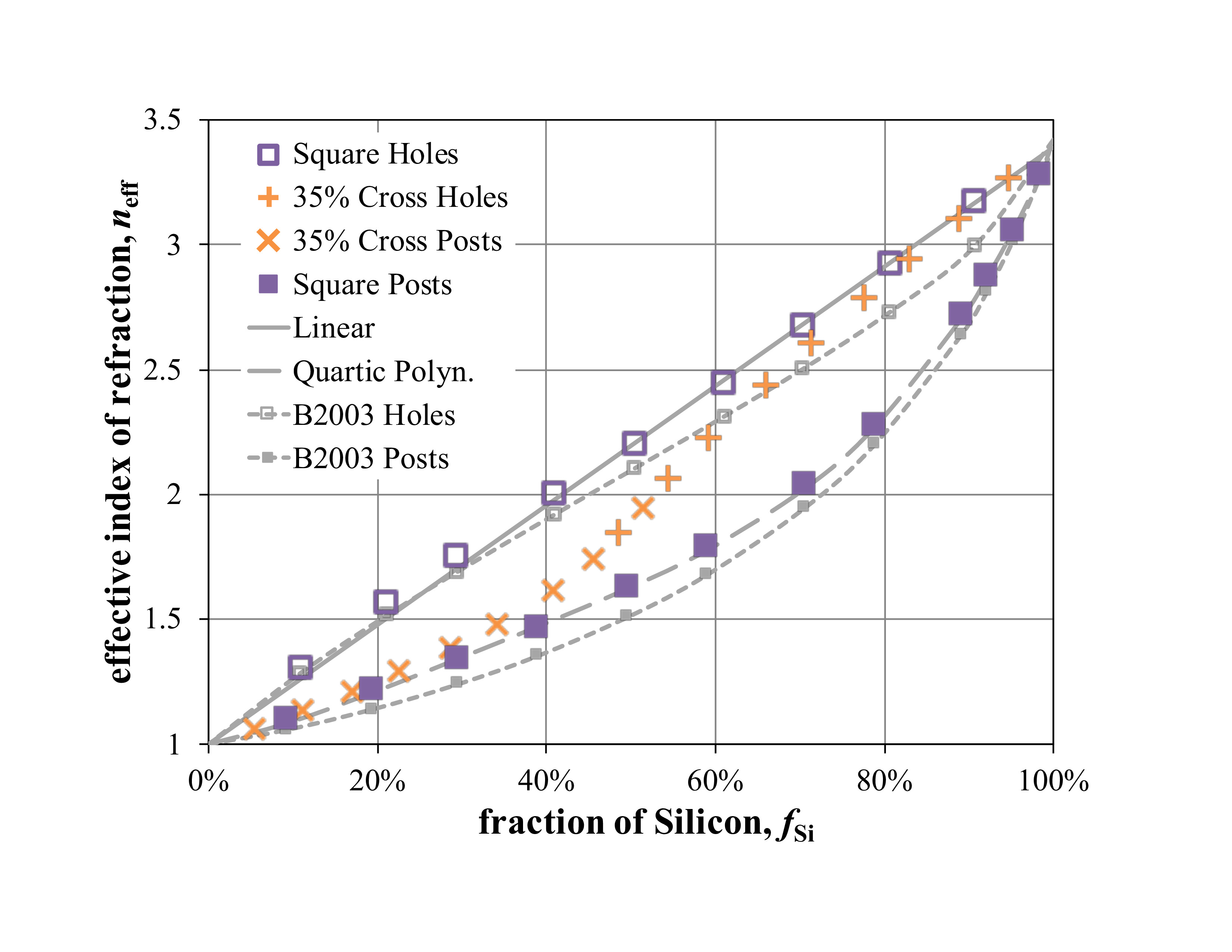}
	\caption{Effective refractive index, $n_{\rm eff}$, of microstructured silicon determined from HFSS calculations. The feature geometries, square holes, cross holes, cross posts, and square posts, all in a square grid, are parameterized by the fill factor, $f_{\rm Si}$. The widths of the cross arms are 35\% of the extent of the enclosing square. Several models are shown for comparison: linear interpolation for holes, a quartic polynomial for posts, and an effective capacitor model for both holes and posts \cite{biber:2003}.  
    }
    \label{fig:neff}
\end{figure}

The results (Fig.~\ref{fig:neff}) show, as expected, the effective index is smaller when there is less silicon, i.e., when $f_{\rm Si}$ is smaller. Holes and posts, however, are significantly different. Posts have a lower effective index than holes when the fraction of silicon is the same. Crosses, formed by indenting the corners of square features, are intermediate between square holes and square posts.  The aspect ratio of their arms and the fill factor both determine their effective indices~\cite{Connors:17}.
For holes, a simple linear model $(n_{\rm eff} - 1) = f_{\rm Si} (n_{\rm Si} - 1) $ provides a reasonable design guide, although it slightly underpredicts the index when the fill factor is small, $f_{\rm Si} < 30$\%. 
An effective capacitor model \cite{biber:2003,Gallardo:17,Chen:14} systematically underestimates the effective index for $f_{\rm Si} > 30$\%.
For posts, an effective capacitor model \cite{biber:2003} provides a general guide but systematically underestimates the effective index, possibly because this static theory does not include high-frequency effects. 
For convenience, we fit our results for square posts with a quartic polynomial: 
($n_{\rm eff} -1) = 4.9 f_{\rm Si}^4 - 6.28 f_{\rm Si}^3 +3.11 f_{\rm Si}^2 +0.66 f_{\rm Si}$. 
Although this fit is only strictly applicable for square posts with the cell size and the frequency range we simulated, it is nevertheless a useful design guide.

Three effects complicate the relation between microstructure geometry and effective index. First, the effective thickness of the microstructured layer differs by a few percent from the physical thickness. The magnitude and sign of this difference depend on both the feature geometry and the fill factor. Second, because the wavelength-to-grid ratio is frequency dependent, the effective index and thickness are also frequency dependent. The magnitude and sign of this chromaticity also depend on both the feature geometry and the fill factor. For holes, we found a positive gradient in the effective index, $\approx 0.5$--$1.5 \times 10^{-4}$~GHz$^{-1}$, while posts show a {\it negative} gradient of similar magnitude. Third, in a multi-layer structure, there will be interactions at the layer interfaces that are not captured by modeling each layer in isolation. 
As a result, any wide-bandwidth, multi-layer AR design must be simulated as a whole to determine its performance. 

The difference between holes and posts is of great practical importance when choosing the geometry to use for each layer of the AR structure.
For an effective index near that of bulk silicon, the fill factor is close to unity and thus the aspect ratio of the removed material is critical to how feasible the structure is: it is difficult to achieve aspect ratios larger than 30:1.  Holes are thus generally easier to physically realize for such high indices because the fill factor at a desired index is lower than for posts. Conversely, an effective index near unity necessitates low fill factor.  Because one must remove so much material, the etching aspect ratio is no longer challenging in general.  Instead, the concerns are now the robustness of the transversely thin structures left by etching and the fractional inaccuracy in their transverse dimensions due to etching tolerances (in dimension and in verticality).  For this case, posts are the better choice because one must leave more material behind to realize a given refractive index, making the design more robust and less sensitive to fabrication imperfections.  For a given fill factor, crosses have effective index values intermediate between those of square holes and posts.  Given the geometries of other layers, crosses may thus provide a means to obtain a desired effective index that is less demanding of the fabrication process than holes or posts.

Additionally, one must consider the relative geometry of posts and holes on different layers, and the considerations depend on the manufacturing technique.  Etching the entire multi-layer structure from the vacuum side alone would require the transverse dimensions of the retained material on a given layer be smaller than the corresponding dimensions of lower layers (greater $n_{\rm eff}$) and larger than those of a higher layers (smaller $n_{\rm eff}$): i.e., like a wedding cake.  This constraint has the greatest impact at the point where the design transitions from posts to holes, which is where $n_{\rm eff} \approx n_{\rm Si}^{-1/2}$: it may not always be possible for the post width to be compatible with the wall thickness of the immediately underlying hole layer.  (Though, for square structures, rotations of one layer by 45\textdegree\  can extend the regime of compatibility (as we do below).)  By wafer-bonding etched structures~\cite{Makitsubo:17}, one can circumvent this constraint by fabricating subsets of the layers on individual wafers using etching from both sides followed by wafer-bonding the etched wafers together, as we plan to do for our proposed four-layer design in Section~\ref{S:next}. 

The above discussion highlights the flexibility of our approach.  For example, in contrast, the dicing saw technique can only produce post structures.
Furthermore, any technique must use wafer bonding to circumvent the above issue of layer-to-layer dimensional compatibility.  Relative to dicing saw and laser techniques, etching permits use of the thinnest wafers in concert with wafer bonding.

\subsection{Microstructure Design and Refinement}
\label{sS:design_2layer_hfss}

\begin{table}[htbp]
\centering
\caption{\bf Antireflection structure design parameters}
\begin{threeparttable}
\begin{tabular}{ccccc}
\hline
Layers & $n_{\rm AR}$ & $t$ [$\mu$m] & Shape & $s$ [$\mu$m] \\
\hline
one & 1.85 & 162  & post & \phantom{0}99 \\
one & 1.85 & 162  & hole & 101 \\
\hline
\multirow{2}{*}{two} & 1.39 & 216 & rotated post & \phantom{0}72 \\
 & 2.46 & 122 & hole & \phantom{0}77 \\
\hline
\end{tabular}
We fabricated three AR treatments for silicon ($n_{\rm Si} = 3.42$): a single layer of posts, a single layer of holes, and a layer of rotated posts above a layer of holes. All features are squares of size $s$ in a square grid with $\Lambda = 125$ $\mu$m spacing. All layer thicknesses, $t$, are one quarter wavelength at the target bandpass's center frequency $\nu_0 = 250$ GHz. 
\end{threeparttable}
\label{tab:AR_design}
\end{table}

\begin{figure}[ht]
    \centering
    \includegraphics[width=\linewidth]{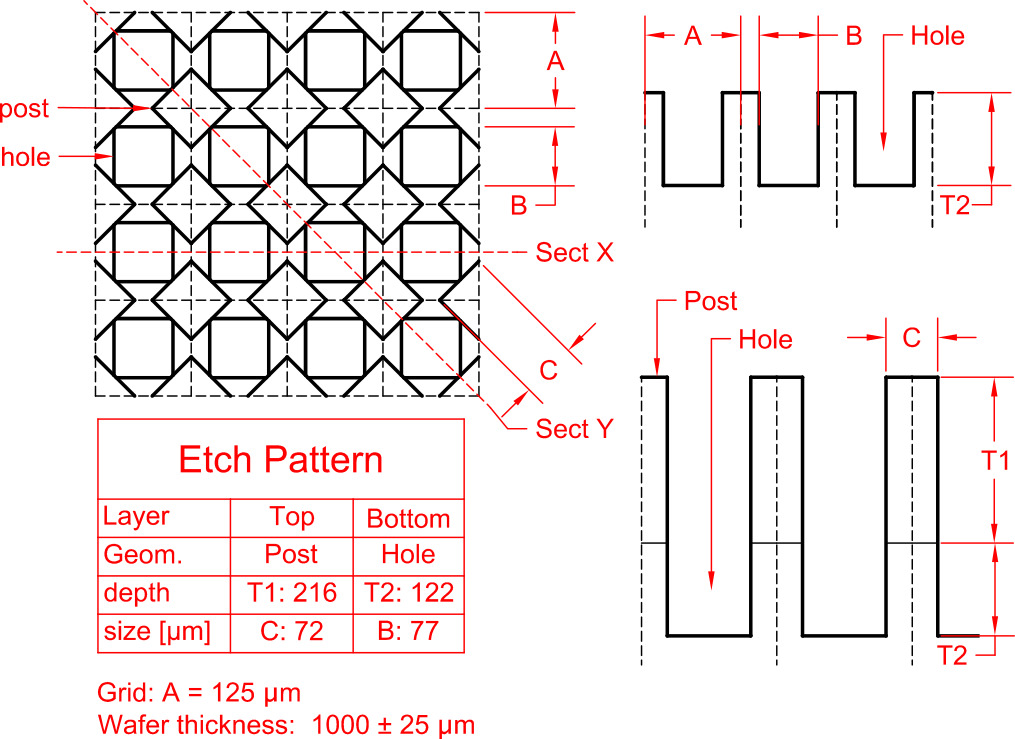}
	\caption{Schematic of the hole and post structures for the two-layer AR structure, neglecting the intrusions in the post walls and other fabrication nonidealities.}
    \label{fig:2lay_schem}
\end{figure}

\begin{figure}[ht]
    \centering
    \includegraphics[width=0.8\linewidth]{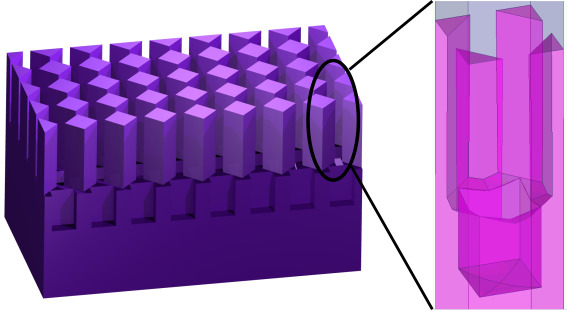}
	\caption{(Left) Three-dimensional model of the two-layer AR structure shown in Fig.~\ref{fig:2lay_schem}.  (Right) HFSS periodic cell used for the simulations, now incorporating tapering of vertical walls and cupping of the bottoms of etched features characteristic of the DRIE process (Section~\ref{sS:Tolerancing}).  We do not show the filleted corners or the intrusions into the posts due to the hole-etching step.}
    \label{fig:2lay_3D}
\end{figure}

\begin{figure}[ht!]
  \includegraphics[width=0.85\linewidth]{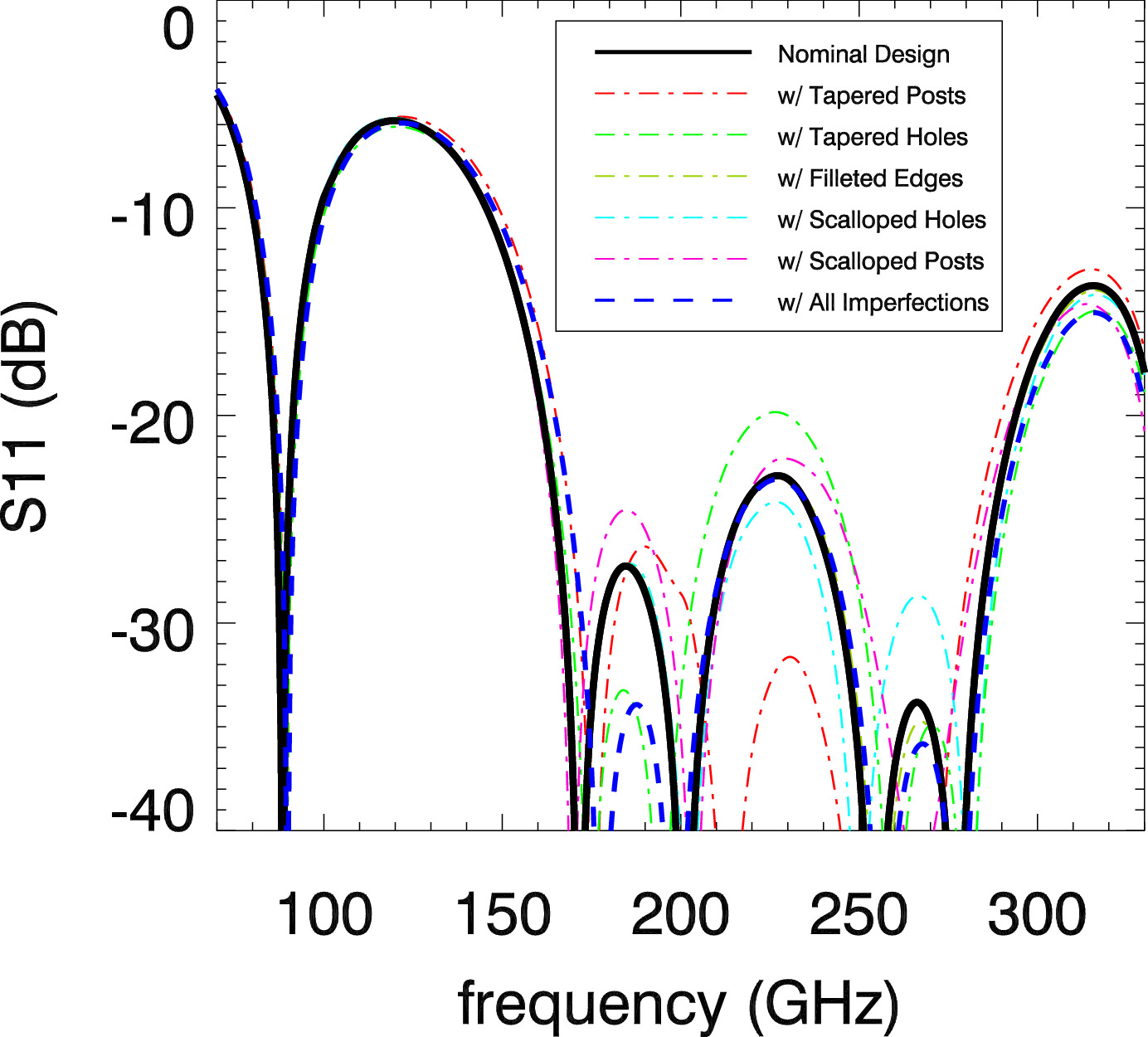}
  \caption{HFSS calculations of our nominal two-layer AR design and of the impact of fabrication nonidealities on its performance.  We describe the modeled nonidealities in Section~\ref{sS:Tolerancing}.  We show the simulated performance of the idealized structure (solid black), of the structure with any one of the nonidealities included (dot-dashed), and of the structure with all of the nonidealities included (dashed blue). }
\label{fig:tolerance}
\end{figure}

Combining our AR designs (Section~\ref{sS:AR_design}) with our calculations of the effective index of microstructures (Section~\ref{sS:design_neff_hfss}), we designed both one- and two-layer AR treatments for fabrication. We directly applied the effective index calculations to obtain the one-layer designs (Table \ref{tab:AR_design}). Although their fill factors are quite different, both holes and posts have similar transverse dimensions and both are equally straightforward to fabricate.

For the two-layer design, we applied the aforementioned considerations about the relative ease of fabricating holes and posts to focus on a design consisting of a layer of square posts above a layer of square holes with dimensions as given in Table~\ref{tab:AR_design}.  We situated the posts above the intersections of the walls of the holes in order to fabricate the structure by multi-depth etching from the vacuum side alone.  Even then, the posts intruded into the holes slightly.  We minimized this intrusion by rotating the posts by 45\textdegree.  See Fig.~\ref{fig:2lay_schem} and Fig.~\ref{fig:2lay_3D}.  For HFSS modeling, we allowed the 72~$\mu$m width of the posts to extend beyond the available diagonal dimension of 67.9~$\mu$m, resulting in a 2.9~$\mu$m chamfer of the 77~$\mu$m holes at each corner.   In practice, we allowed the hole etch pattern to act on the post layer, resulting in right-triangular intrusions of the holes into the post walls with side length 2.9~$\mu$m and hypotenuse 4.1~$\mu$m.  The good match of the HFSS calculations to the data in Section~\ref{S:2lay} confirms these differences do not yield discrepancies above $-$$20$~dB reflectance.  
For completeness, we also modeled designs consisting of two layers of square holes and two layers of square posts, and we found all three designs yielded similar behavior, though with some variation in the heights of the reflectance maxima.  

\begin{figure*}[ht]
    \centering
  	\includegraphics[width=0.73\linewidth]{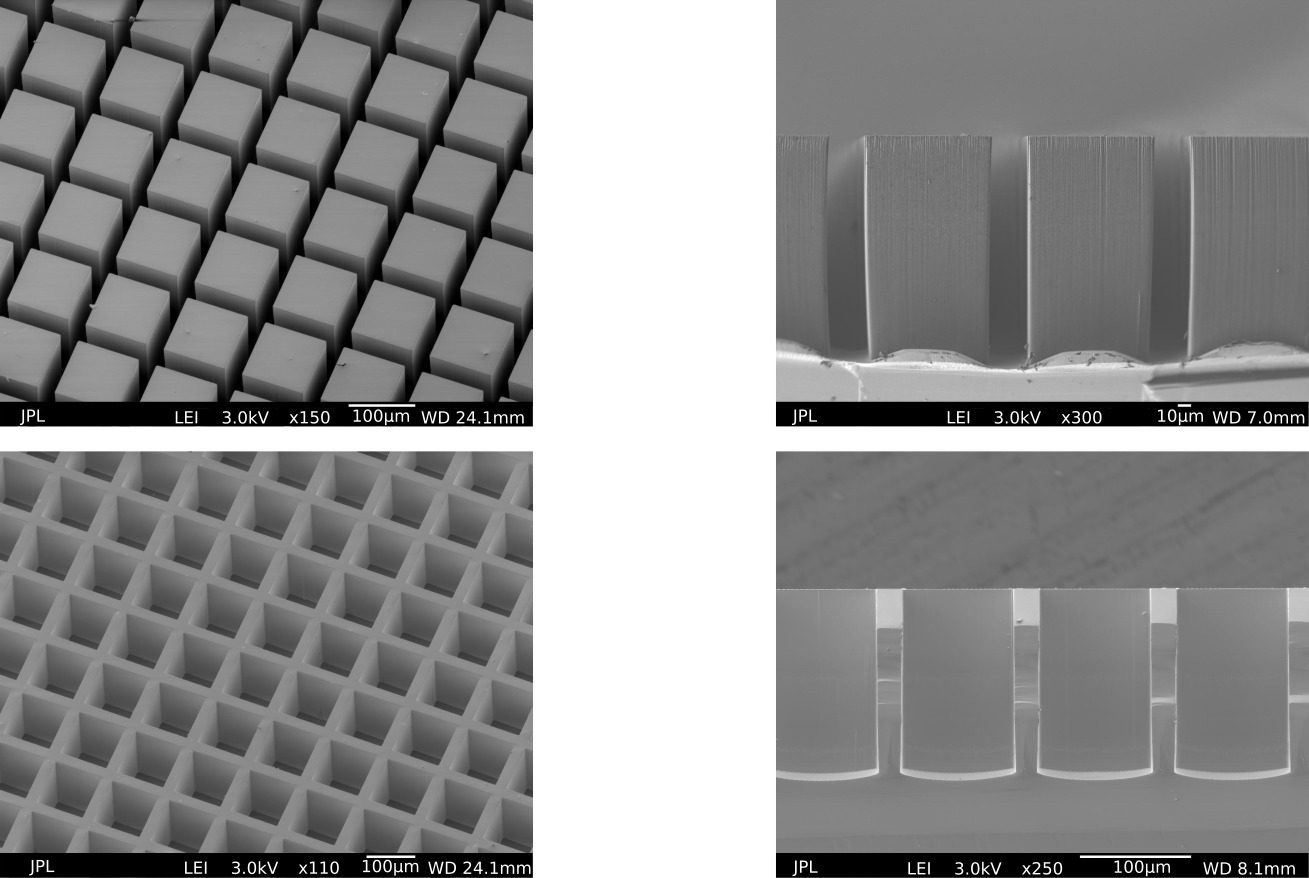}
    \caption{SEM images of one-layer AR structures. (Top) Square posts. (Bottom) Square holes. (Left) Isometric view. (Right) Cross-section view of cleaved samples.}
    \label{fig:sem_1layer} 
\end{figure*}

We did not further optimize this design for minimum in-band reflectance or to strictly satisfy the desired 190--310~GHz band definition prior to manufacture.  The HFSS analysis of a flat, 1~mm thick wafer with the structure on both sides (Fig.~\ref{fig:tolerance}) indicated the passband of the structure would be shifted to slightly lower frequency than desired, $<$$-$20~dB over 160--280~GHz.  Rather than refine the design to precisely match the desired band, we deemed it more important to demonstrate the structure could be reliably fabricated and accurately modeled.  The shift may be due to interaction between the two layers.



\subsection{Effects of Fabrication Nonidealities}
\label{sS:Tolerancing}

Although our DRIE process (Section~\ref{S:Fabrication}) allows us to fabricate patterns that closely match our idealized rectilinear designs, SEM images (Fig.~\ref{fig:sem_1layer} and Fig.~\ref{fig:sem_2layer}) show the fabricated patterns do not perfectly reproduce the designed geometry: the bottoms of the holes, along with the spaces between the posts, show a cupped profile; the transverse dimension of the etch tends to expand with depth, resulting in posts that are slightly narrower at the bottom compared to the top and holes that are slightly wider at the bottom compared to the top; and, none of the edges are perfectly sharp, but instead have a slightly rounded, or filleted, profile. 
  
To quantify the impact of these fabrication nonidealities on the performance of our two-layer AR design, we performed calculations using HFSS at normal incidence, assuming a 1~mm total wafer thickness (including the etched layers on both sides, as fabricated (Section~\ref{S:Fabrication})).  We included the above nonidealities based on their typical sizes as measured by SEM: we approximated the cupped profiles at the bottom of the etched volumes using linear pyramidal shapes with 10~$\mu$m heights; we modeled the horizontal dimensions of the posts/holes with linear profiles that expanded by 4~$\mu$m from top to bottom; and, we added a 2~$\mu$m radius fillet to all sharp edges and corners.  We show the results in Fig.~\ref{fig:tolerance}.  The nominal design provides $<$$-$20~dB reflectance over the band 160--293~GHz.  The largest effects are: tapering the holes causes one Fabry-P\'erot fringe to rise a small fraction of a dB above $-$20~dB; and, tapering the posts causes the band to narrow slightly, shifting the band edges inward by 2--3~GHz.  When we include all the nonidealities, the effects partially cancel, with the most important result being a slight upward shift of the band edges by 1--2~GHz.  Overall, these simulations indicated that the modeled nonidealities had noticeable but acceptably small effects on performance.

\section{Fabrication}
\label{S:Fabrication}

\subsection{Substrates}

We used 100 mm diameter, 1 mm thick wafers $<$100$>$ wafers, optically polished on both sides and specified to have bow/warp $<$30~$\mu$m, total thickness variation $<$5~$\mu$m, and no more than 10 particles above 0.3~$\mu$m in size per face.  The material is high-resistivity float-zone silicon, lightly n-type doped with phosphorus, with $>$10~k$\Omega$~cm resistivity.  The implied loss tangent should be $<$$7\,\times\,10^{-5}$ at 250~GHz and room temperature, resulting in a loss of $<$$0.1$\% over the 1~mm thickness (with $\tan \delta = \left( 2 \pi \nu \epsilon_0 \epsilon_r \rho\right)^{-1}$ where $\nu$ is the frequency, $\epsilon_0$ the permittivity of vacuum, $\epsilon_r$ the relative permittivity of silicon, and $\rho$ its resistivity).  Though not immediately relevant here, we note for completeness that the specification on minority carrier lifetime is $>$1~msec.

\subsection{Antireflection Structure Fabrication}

\begin{figure*}[t!]
    \centering
	\includegraphics[width=0.95\linewidth]{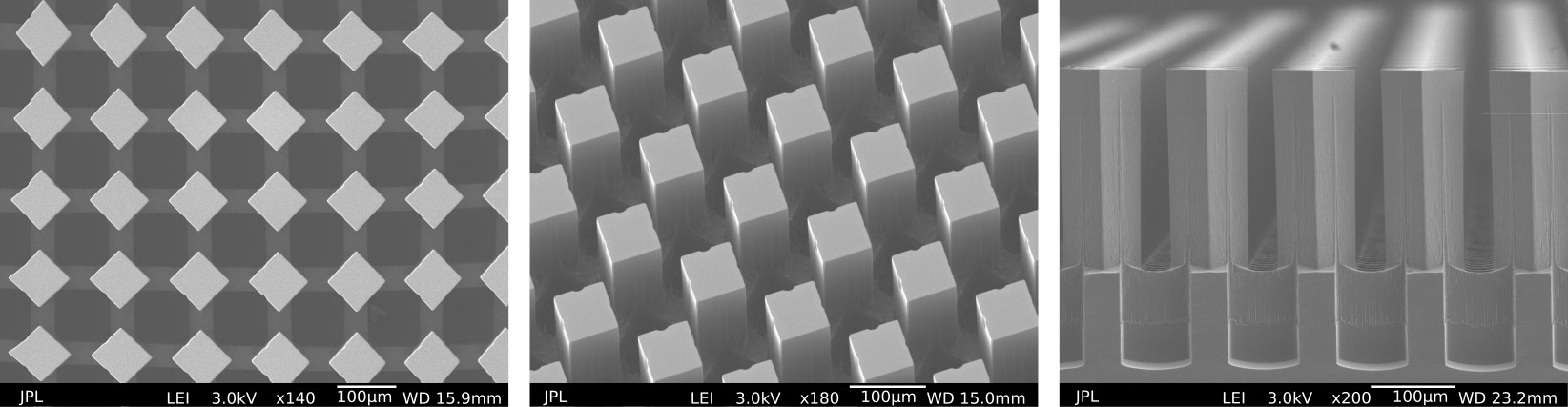}
    \caption{SEM images of two-layer AR structure.  (Left) Top view. (Center) Isometric view. (Right) Cross-section of cleaved sample.}
    \label{fig:sem_2layer}
\end{figure*}

Deep reactive-ion etching (DRIE) is a dry etching technique, relying on plasma-etching of bulk silicon.  We employed the Bosch technique because it is known to produce nearly vertical sidewalls, high aspect-ratio features, and is compatible with mass production~\cite{Chattopadhyay:17}. It utilizes SF$_6$ and C$_4$F$_8$ as the main gases, with alternating etching and passivation steps~\cite{Laermer:1996}.  It is compatible with various mask materials, including photoresist, silicon oxide/nitride, or metals.  In all cases, the desired etch depth divided by the DRIE process selectivity (the ratio of silicon to mask material etch rates) determines the mask thickness.

For multi-layer AR structures, we began with the multi-step DRIE process previously reported in~\cite{Jung-Kubiak:16}, where the SiO$_2$ mask was patterned with steps of different thicknesses, each in proportion to the desired etch depths of the silicon layers.  We made one modification for this work, using a photoresist mask for the etch of the highest-index layer (closest to bulk silicon), due to differences in process details.  In particular, we found that the DRIE process has poorer selectivity between SiO$_2$ and silicon than in~\cite{Jung-Kubiak:16}.  This may be due to the larger volumes of silicon being removed.  This observation, along with the larger etch depths used here (338~$\mu$m total etch depth here, 254~$\mu$m in~\cite{Jung-Kubiak:16}), would have necessitated an impractically thick SiO$_2$ mask for the last etch step, motivating the use of photoresist instead.  That is, we used a single photoresist mask on top of a single SiO$_2$ mask to fabricate our two-layer design. 
(For one-layer structures, we used only a SiO$_2$ mask.)  We grew the thick ($\sim$2~$\mu$m) SiO$_2$ mask under water vapor at 1050~{\textdegree}C directly on the silicon wafers.  We patterned the SiO$_2$ into a mask using an Inductively Coupled Plasma (ICP) machine with O$_2$ and CHF$_3$ gases and an etch rate of about 250~nm~min$^{-1}$.  We used conventional UV photolithography to expose the photoresist mask, which was spun on after the SiO$_2$ was patterned.  

We etched the silicon with a Plasma-Therm VERSALINE Deep Silicon Etcher, using a modified 3-step Bosch process: passivation step, etch A step, etch B step. In addition to traditional SF$_6$ and C$_4$F$_8$ gases, we added Ar to each of the three steps to keep the plasma stable during the short transitions.  To provide smooth surfaces and vertical sidewalls for even the large aspect-ratio features, and to remove the large amounts of silicon demanded by our design, we optimized the gas ratio, step timing, and power levels.  
We performed the etches at a chamber pressure of 20--35~mTorr, gas flow rates of 100--150 sccm for SF$_6$ and C$_4$F$_8$ and 30 sccm for Ar, inductively coupled plasma RF power of 1500~W, and a chuck temperature of 15~{\textdegree}C.  We also bias the substrate with RF power (we do not supply this power value because it is very machine-dependent).  Each step in the process is a few seconds long, with specific lengths depending on the desired etch depth and calibration from test wafers.
The first DRIE step, using the photoresist mask, nominally provided an etch depth of 122~$\mu$m to produce the hole features of the deepest AR layer (but see below for a correction to this depth).  The photoresist mask protected the post regions as well as the walls between the holes.  After stripping the photoresist, the second DRIE step etched a depth of 216 $\mu$m more, bringing both patterns to their target depths.  The SiO$_2$ mask protected the posts during this step.  

\begin{figure}[t!]
    \centering
    \includegraphics[width=0.8\linewidth]{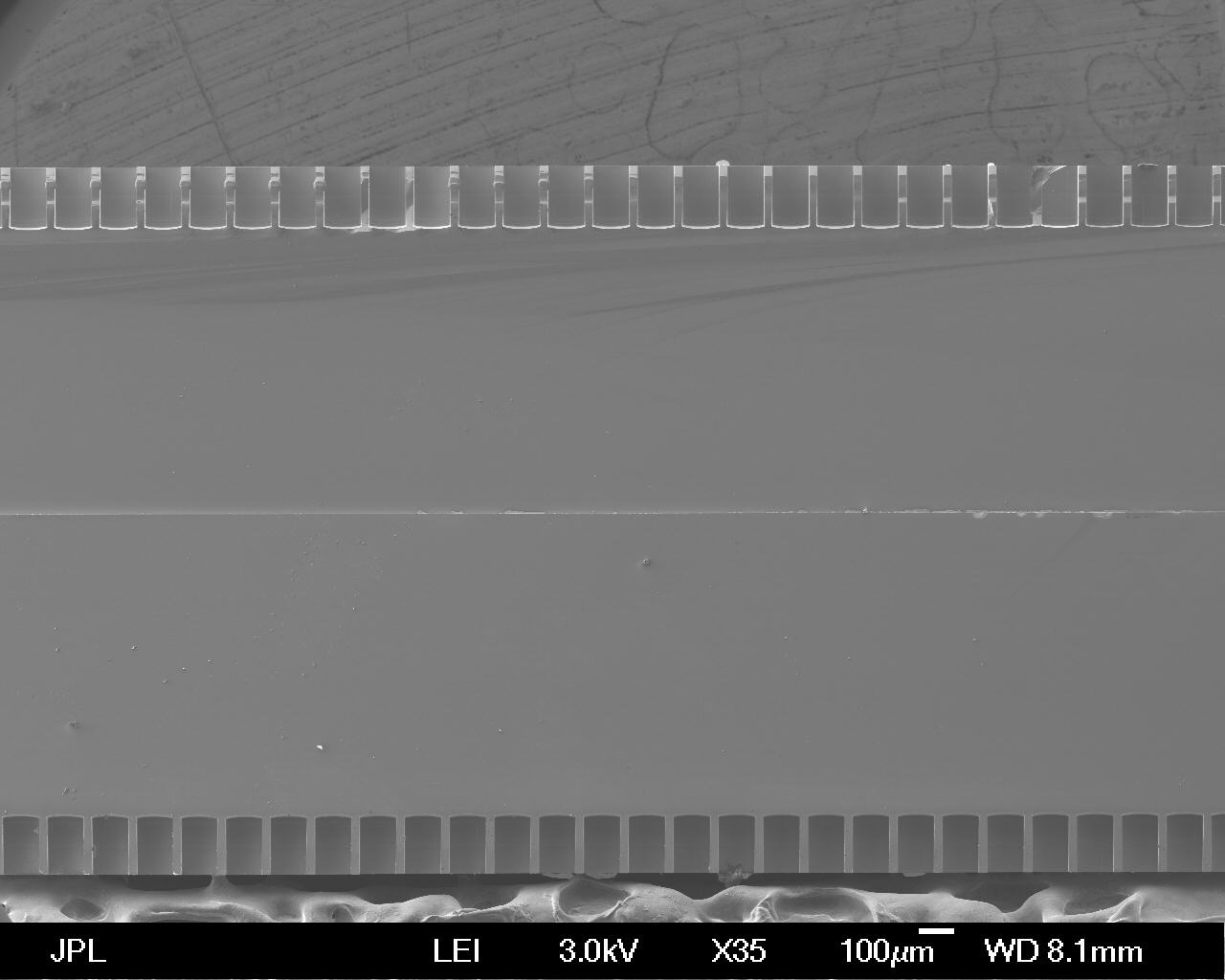}
	\caption{SEM of boundary between bonded wafers with hole AR structures.  No gaps are visible at the boundary at this resolution, though some imperfections are present.}
    \label{fig:waferbond}
\end{figure}

During initial testing, we found that, during the second etch step, which both creates the post layer and extends the holes down to their final depth, the etch rate of the deeper hole layer, relative to that of the post layer, slowed down dramatically with time due to its depth in the silicon.  To overcome this issue, we over-etched to 190~$\mu$m (as opposed to 122~$\mu$m) during the first etch step.  With this change, the holes reached their final target depth of 338~$\mu$m (= 122 + 216~$\mu$m) as the post layer reached its final 216~$\mu$m depth target.
Additionally, to compensate for the slight undercut during DRIE, we augmented the feature dimensions on the photolithographic mask by 2--3~$\mu$m on each side relative to the design, with the corrections determined empirically by etching pathfinder wafers.  Overall, we found the first step to have an approximate silicon-to-photoresist selectivity of 25:1 for an etching time of 56~min and the second step to have a silicon-to-SiO$_2$ selectivity of 140:1 for an etching time of 68~min. 

After etching, the remaining SiO$_2$ mask was removed with hydrofluoric acid and the wafer was cleaned with an O$_2$ plasma for 1~hr at 1000~W.
We also performed a short thermal oxidation and etching step to improve the etched surfaces' morphology and smoothness~\cite{Reck:2014}.   We grew a sacrificial layer of SiO$_2$ in an oxidation furnace at 1050~{\textdegree}C using water vapor for approximately 1 hr, which we then removed with the above recipe.

After DRIE, we verified the depths by scanning electron microscopy (SEM), shown in Fig.~\ref{fig:sem_1layer} and Fig.~\ref{fig:sem_2layer}.  We demonstrated good control of all lateral dimensions (A, B and C from Fig.~\ref{fig:2lay_schem}). 

\begin{figure*}[ht]
    \centering
    \includegraphics[width=0.91\linewidth]{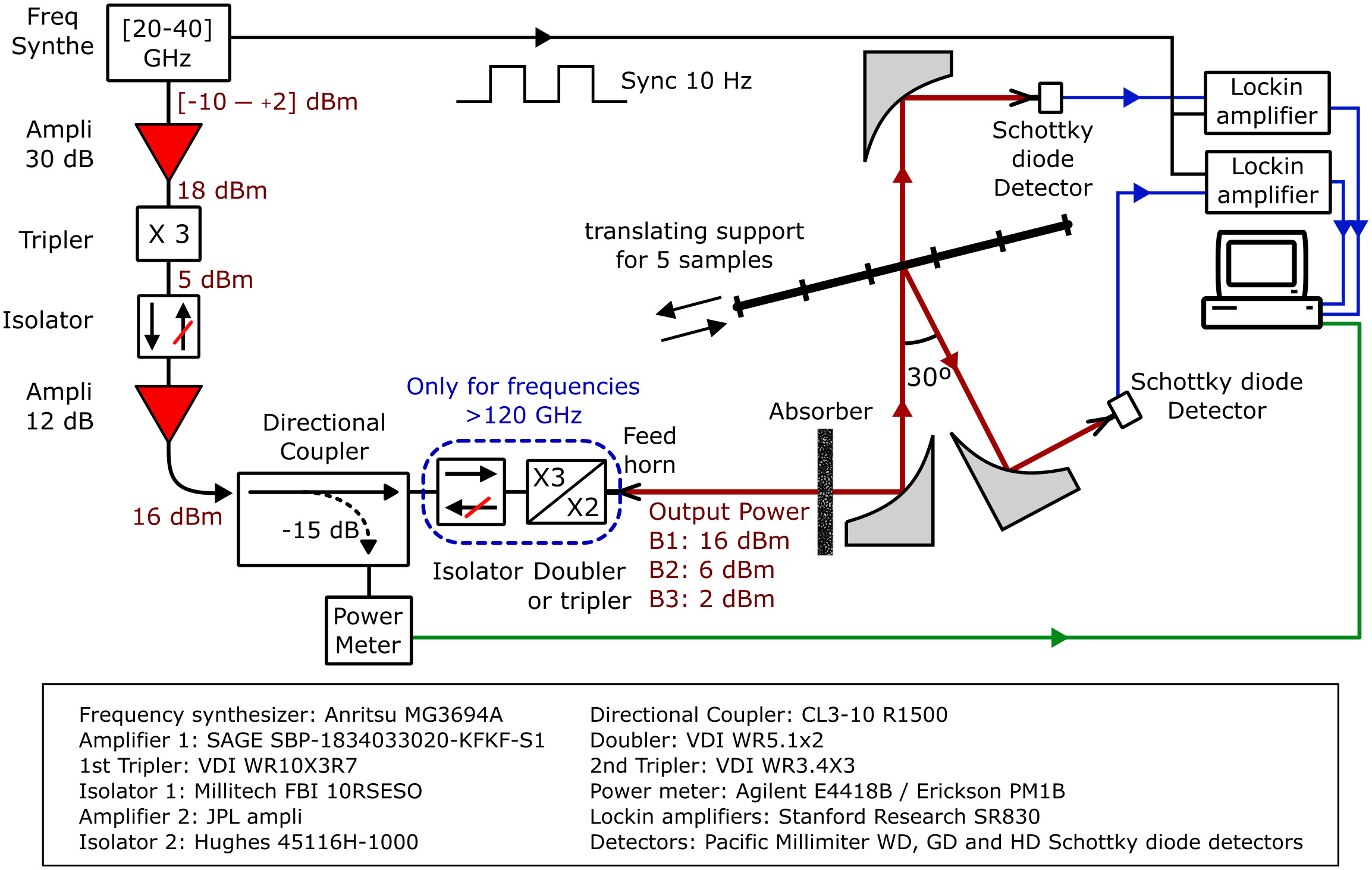}
	\caption{Schematic of the test bench.}
    \label{fig:schematic}
\end{figure*}

\subsection{Wafer Bonding}
\label{sS:waferbonding}



As explained in Section~\ref{sS:approach}, wafer-bonding of patterned silicon wafers is integral to our approach for constructing broadband, antireflection-textured, gradient-index silicon optics.  We employ a standard process, ``hydrophilic fusion bonding'' (also known as ``hydrophilic direct bonding'')~\cite{Gosele:98,Tong:99}, which has been used in prior work on one-layer AR structures~\cite{Gallardo:17}. 

Prior to bonding, we performed cleaning and oxidation steps to ensure a high-quality bond.  We began by using a 1:1 mixture of sulfuric acid (H$_2$SO$_4$) and hydrogen peroxide (H$_2$O$_2$) (also known as Piranha solution) for about 10 minutes to remove organic residues, followed by a long rinse under water.   We followed this with a 1~hr O$_2$ plasma clean.  We then grew a thin ($\sim$500~nm) layer of SiO$_2$ via a 1~hr exposure to water vapor in an oxidation furnace at 1050~{\textdegree}C with a nitrogen atmosphere.  Finally, immediately before the initiation of the bond, we performed a two-step cleaning process with solutions of RCA-1 (H$_2$O/H$_2$O$_2$/NH$_4$OH (5:1:1)) heated to 80~{\textdegree}C and RCA-2  (H$_2$O/H$_2$O$_2$/HCl (4:1:1)) heated to 70~{\textdegree}C to remove any remaining organic and/or metallic particles.  

Next, to initiate bonding, we aligned the two silicon wafers using a contact mask aligner with backside capabilities and brought them into contact, forming van der Waals bonds between the wafers.  We allowed the wafers to rest for at least 12 hours at room temperature to provide time for any bubbles to dissipate and to release surface tension between the two wafers.  We strengthened the still temporary bond by bringing the two wafers to 450~{\textdegree}C for at least 12 hours.  After this step, we inspected the wafer pair under an IR microscope for any remaining bubbles or voids.  Unsatisfactory inspection results would have led us to separate the wafers and restart the process, but we had no such failures during bonding of the sample reported on in Section~\ref{S:waferbonding_results}.  Finally, we proceeded to create a chemical bond~\cite{Suni:06} between the wafers by heating them to 1050~{\textdegree}C.   Since the quality of the bond increases for longer annealing times, we annealed our wafers for 1--3 days in a dry environment (no water vapor).  The thermal oxide layer grew to a thickness of 0.4--1~$\mu$m during this step.  We show a cross-section of a bonded wafer in Fig.~\ref{fig:waferbond}.

\section{Measurement Setup}
\label{S:test_setup}

We used a scalar spectrometer (Fig.~\ref{fig:schematic}) to measure the reflectance and transmittance of samples between 75~GHz and 330~GHz.  The signal generation chain begins with a frequency synthesizer outputting a 20--40~GHz signal that is amplitude modulated (10~Hz for the data shown here).  We amplify and triple this signal, with the tripler having sufficiently high output power between 75 and 115~GHz.  We follow the tripler with an isolator to mitigate standing waves, a W-band amplifier, a directional coupler to provide a monitor signal, and finally a rectangular feedhorn for measurements in the 75--115~GHz band (designated as ``Band 1'').  We use a lookup table based on the measurements of the monitor signal to power-level the W-band signal and, for the higher-frequency bands, protect the doubler/tripler.  The output signal is polarized normal to the optical bench and hence, relative to the sample, normal to the plane of incidence or TE.  For measurements at higher frequencies, we employ another isolator, either a doubler (to reach ``Band 2'': 140--220~GHz) or a tripler (to reach ``Band 3'': 220--330~GHz), and a band-appropriate rectangular feedhorn.  Fig.~\ref{fig:schematic} shows, in dark red, the power levels at various points in the multiplier chain as well as the band-specific  power at the output of the chain.  Fig.~\ref{fig:schematic} also lists identifying information for each element of the setup.

\begin{figure}[t]
    \centering
    \includegraphics[width=\linewidth]{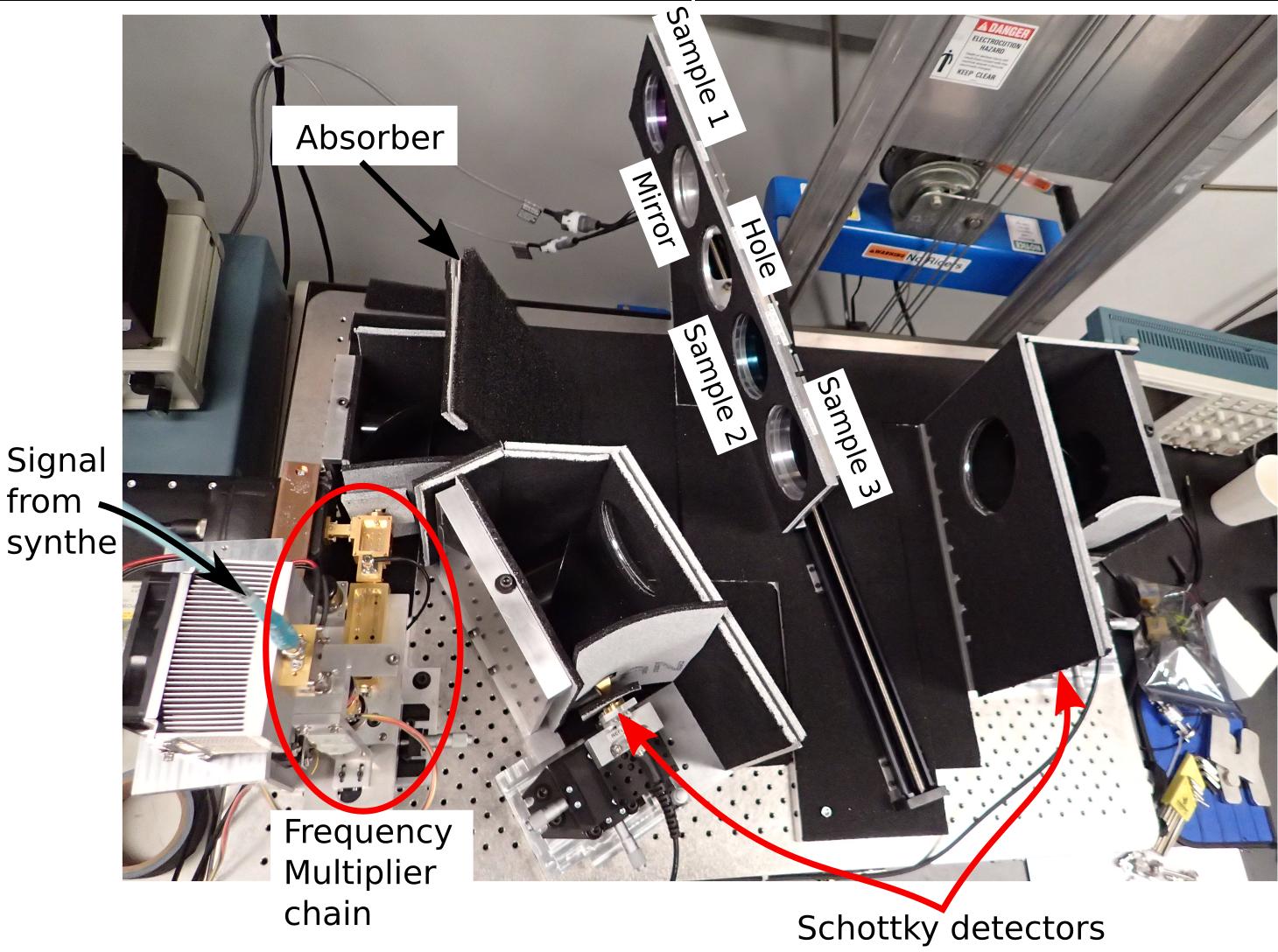}
	\caption{Top view of the test setup.}
    \label{fig:pict_setup}
\end{figure}

We attenuate the signal emitted by the generation chain with an Eccosorb\textsuperscript{\textregistered} HR-10 foam absorber (attenuation increasing from $-$10~dB in Band 1 to about $-$20~dB in Band 3) to mitigate standing waves.  A parabolic mirror then collimates the beam and directs it to the sample.  We mount the samples to be measured in a translating support that makes a 15\textdegree\ angle with the incident beam.  The support has three measurement locations for samples and two calibrator locations, one used for a mirror and the other with no element (for full transmission).  To measure the transmission of a sample, we calculate the ratio of the power transmitted by the full transmission calibrator and by the sample, while the reflectance is the ratio of the power reflected by the sample and by the mirror.  For both the reflection and transmission arms, we focus the signal from the sample via a parabolic mirror to a feedhorn coupled to a Schottky diode power detector.  We monitor the diode voltages with lockin amplifiers.  We use band-appropriate feedhorns and diode detectors.  To reduce unwanted reflections, we cover all flat surfaces with the same foam absorber as used above (see Fig.~\ref{fig:pict_setup}).

\begin{figure}[ht]
    \centering
    \includegraphics[width=\linewidth]{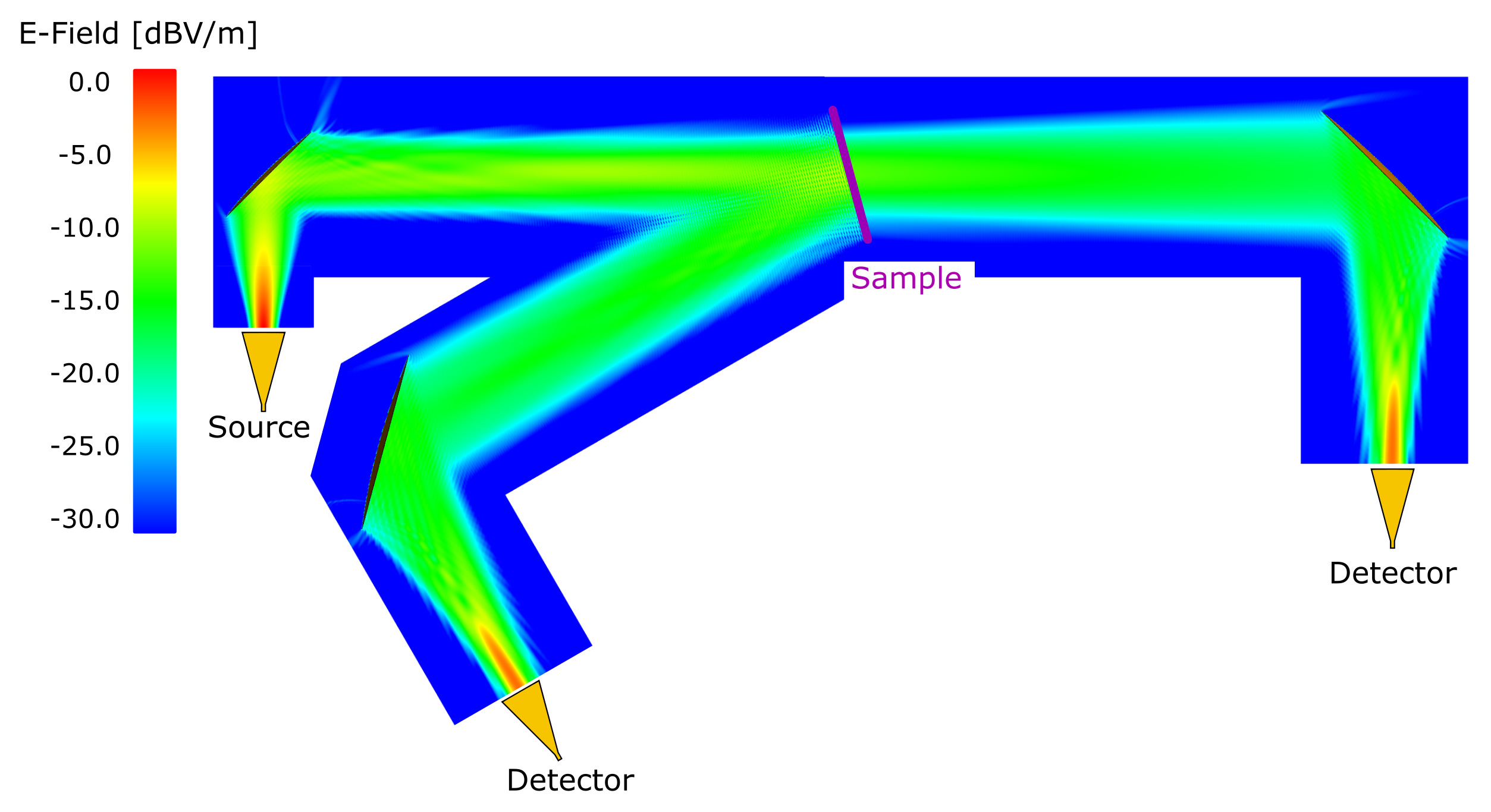}
	\caption{Propagation of the Gaussian beam along the optical path of the test setup at 100~GHz (Band 1) using Feko\textsuperscript{\textregistered}. This figure is the combination of the near-field patterns given by two simulations, one with no sample and one with a perfectly reflective sample.  See text for details.}
    \label{fig:optical_path}
\end{figure}

In order to accurately evaluate the intensity distribution and wavefront of the beam along the optical path, especially at the sample and at the receiving horn positions, we used Feko\textsuperscript{\textregistered} to simulate the propagation of the beam (Fig.~\ref{fig:optical_path}). The simulation uses the Multi-Level Fast Multipole Method (MLFMM) solver, which is based on the method of moments.  
In addition to basic Gaussian-beam propagation calculations (which can be performed without the use of electromagnetic simulation software), it is possible to use the real, imperfectly Gaussian shape of the beam emitted by the horn for each of the three bands.  The simulation also accounts for the transformation of the beam by the parabolic mirrors and the slight truncation of the beam by the different elements (mirrors, sample).

The setup suffers modest systematic effects arising from the modification of the beam propagation by the sample.  Because of the 15\textdegree\ incidence angle, a 1~mm thick silicon sample shifts the beam transversely by 0.19~mm.  Additionally, as detailed in \cite{hanna:69,nemoto:88}, when a Gaussian beam passes through a flat dielectric slab (even at normal incidence), the waist position of the transmitted beam differs from that of the incident beam.  For our setup and our typical silicon sample thickness of 1~mm, this effect artificially shortens the distance between the beam waist and the post-sample parabolic mirror (in the transmission arm only) by 0.7~mm.  We used Feko\textsuperscript{\textregistered} to quantitatively evaluate these two effects, finding a 0.4\% reduction in power received at the transmission-arm detector at 100~GHz and 0.5\% at 300~GHz.  For the reflection arm, the effect only occurs for the component of the signal reflected from the backside of the sample, so it modulates an already small reflectance by a few percent relative.  (It would be more important for samples with higher reflectance.)  The sub-1\% systematic error on the transmission measurement was not important for this work, as we relied entirely on the reflectance measurement to quantify the effectiveness of the AR structure specifically for the above reason: multiplicative systematic uncertainties are much less important for the near-zero reflectance.

\begin{figure}[t!]
    \centering
 	\includegraphics[width=0.98\linewidth]{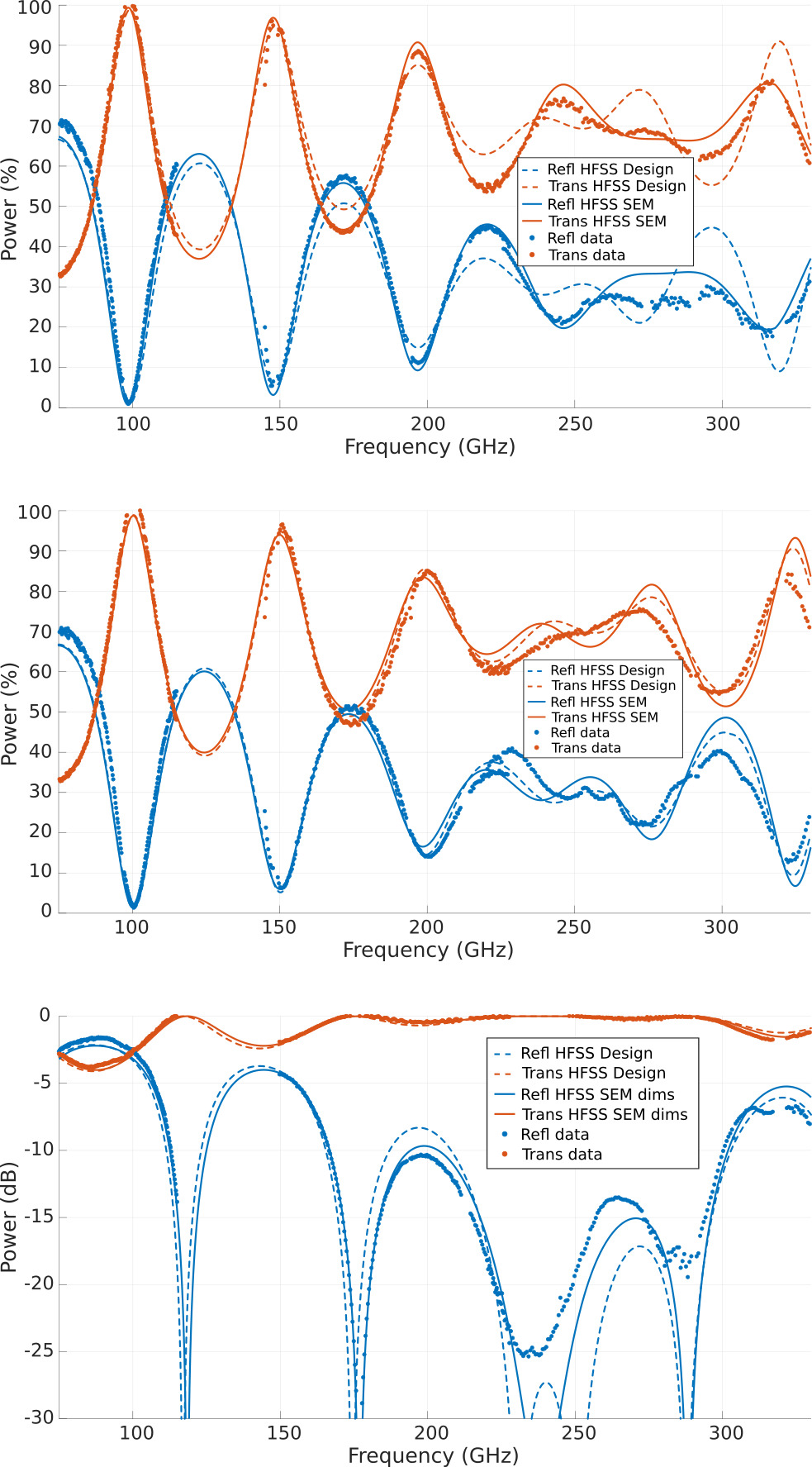}
    \caption{\label{fig:data_1layer}
Reflectance and transmittance measurements of silicon wafers with one-layer AR structures. We compare to the results of finite element calculations using the design (dashed line) and measured (solid line) dimensions (Table~\ref{tab:dims1_lay_sem}) and 15\textdegree\ incidence angle. (Top) A single layer of square posts on one side of a wafer. (Middle) A single layer of square holes on one side of a wafer. (Bottom) A single layer of square holes on both sides of a wafer.  Note the upper two panels use linear vertical scales while the bottom uses a logarithmic vertical scale to emphasize the low reflectance.  We discuss possible explanations of the modest discrepancies between simulation and data in the text.
}
\vspace{0.3cm}
\end{figure}


\begin{figure*}[t]
    \centering
	\includegraphics[width=0.94\linewidth]{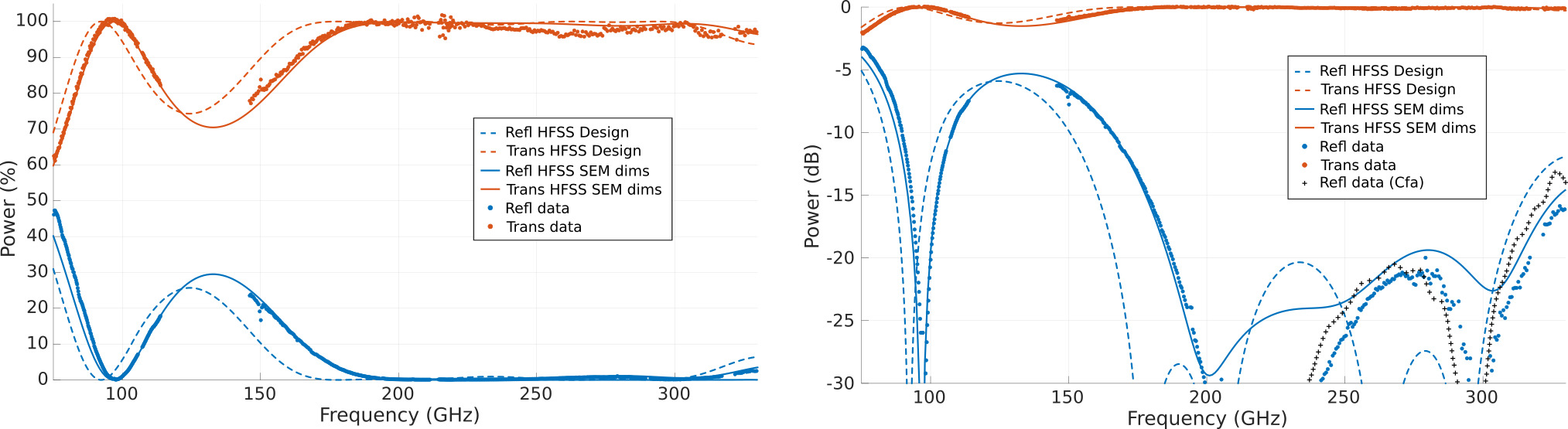}
    \caption{Reflectance and transmittance measurements of a silicon wafer with two-layer AR structures on both sides compared to finite-element simulations. (Left) Linear vertical scale.  (Right) Logarithmic vertical scale.  We show finite-element calculations with the design dimensions (dashed line) and with the dimensions measured from SEM images (solid line), as listed in Table~\ref{tab:dims_2lay_vals}.  Both calculations incorporate the 15\textdegree\ angle of incidence of the measurement as well as the actual wafer thickness (996~$\mu$m).  These changes result in a shift of 3~GHz and some differences in the in-band reflectance relative to the design calculation shown in Fig.~\ref{fig:tolerance}.  The black crosses show the reflectance measurement of the same wafer at normal incidence between 200~GHz and 330~GHz, with an alternate technique, as mentioned in the text.}
    \label{fig:data_2layer}
\end{figure*}

\section{One-Layer Sample Measurements and Comparison to Finite-Element Analysis}
\label{S:1lay_results}

\begin{table}[ht]
  \centering
    \caption{\bf Dimensions of one-layer AR structures}
  \begin{threeparttable}
  \begin{tabular}{lrrrr}
  \hline
    Shape  & \multicolumn{2}{c}{Posts ($n_{\rm eff} = 1.85$)} & \multicolumn{2}{c}{Holes ($n_{\rm eff} = 1.85$)} \\ \hline  
  Dimension         & \begin{tabular}[c]{@{}c@{}}Height \\ $[\mu$m$]$\end{tabular} & \begin{tabular}[c]{@{}c@{}}Width\\ $[\mu$m$]$\end{tabular} & \begin{tabular}[c]{@{}c@{}}Depth\\ $[\mu$m$]$\end{tabular} & \begin{tabular}[c]{@{}c@{}}Width \\ $[\mu$m$]$\end{tabular}
  \\  \hline
  \multicolumn{1}{l}{Design}     & 162       & 99        & 162     & 101        \\ \hline
  \multicolumn{1}{l}{\begin{tabular}[c]{@{}l@{}}Measured \end{tabular}} & 157            & \begin{tabular}[c]{@{}r@{}}top: 97\\ base: 93\end{tabular} & 171          & \begin{tabular}[c]{@{}r@{}}top: 99\\ base: 101\end{tabular}                    \\ \hline
  \end{tabular}
We compare the design and measured (by SEM) dimensions for the two one-layer single sided wafers. The top and base dimensions of the posts and holes are the width at the top of the structures (toward vacuum) and at the base (toward bulk silicon), respectively.  All layers use a 125~$\mu$m cell size (grid spacing).
  \end{threeparttable}
  \label{tab:dims1_lay_sem}
\end{table}

In order to validate the fabrication technique and the test setup, we first fabricated and measured two one-layer AR structures, one with square holes and one with square posts (see Fig.~\ref{fig:sem_1layer}).  We optimized the designs for maximum transmission at 250~GHz, resulting in the dimensions given in Table~\ref{tab:dims1_lay_sem}.   
We fabricated these structures using DRIE as described in Section~\ref{S:Fabrication}.  SEM measurements of a cleaved wafer confirmed that the dimensions of the fabricated structures were close to the design dimensions (within $\pm$6\%).  We tested the samples using the setup described in Section~\ref{S:test_setup}.  We present the reflectance and transmittance results of these one-layer AR coatings in Fig.~\ref{fig:data_1layer}.  Their good agreement with the HFSS calculations validates the one-layer HFSS calculations, the fabrication process, and the testing setup, important steps toward multi-layer AR structures.  It may be possible to explain the modest discrepancies by the same effects we discuss in Section~\ref{S:2lay}.

\section{Two-Layer Sample Measurements and Comparison to Finite-Element Analysis}
\label{S:2lay}

Our two-layer design, presented in Fig.~\ref{fig:2lay_schem} and Fig.~\ref{fig:2lay_3D}, consists of a top layer of square posts ($n_{\rm eff}$ = 1.39) and a bottom layer of square holes ($n_{\rm eff}$ = 2.46), designed for a bandpass centered on 250~GHz.  We used the multi-depth DRIE process described in Section~\ref{S:Fabrication}.  We measured both samples with our test bench, and we cleaved and analyzed one of these samples with a SEM (Fig.~\ref{fig:sem_2layer}) to obtain accurate dimensions.  The two sets of optical measurements were quite consistent, indicating reproducibility of the AR structures.  
We updated the HFSS simulations described in Section~\ref{sS:design_2layer_hfss} using the dimensions derived from the SEM measurements (see Table~\ref{tab:dims_2lay_vals}) so that discrepancies would inform us about differences between theory and measurement, as well as random variations across the wafer, rather than being dominated by the mean differences between the designed and fabricated structures.  

\begin{table}[ht!]
  \centering
  \caption{\bf Dimensions of two-layer AR structure}
  \begin{threeparttable}
  \begin{tabular}{lrrrr}
   \hline
    Shape   & \multicolumn{2}{c}{Posts ($n_{\rm eff} = 1.39$)}  & \multicolumn{2}{c}{Holes ($n_{\rm eff} = 2.46$)} \\ \hline
  Dimension     & \begin{tabular}[c]{@{}c@{}}Height \\ T1 [$\mu$m]\end{tabular} & \begin{tabular}[c]{@{}c@{}}Width\\ C [$\mu$m]\end{tabular} & \begin{tabular}[c]{@{}c@{}}Depth\\ T2 [$\mu$m]\end{tabular} & \begin{tabular}[c]{@{}c@{}}Width \\ B [$\mu$m]\end{tabular} \\ \hline
  Design  & 216   & 72     & 122        & 77      \\ \hline
  \begin{tabular}[c]{@{}l@{}}Measured \\ Face 1\end{tabular} & 228 & \begin{tabular}[c]{@{}l@{}}top: 71\\ base: 66\end{tabular} & 112  & \begin{tabular}[c]{@{}l@{}}top: 82\\ base: 83 \end{tabular}      \\ \hline
  \begin{tabular}[c]{@{}l@{}}Measured \\ Face 2\end{tabular} & 210 & \begin{tabular}[c]{@{}l@{}}top: 71\\ base: 65\end{tabular} & 121 & \begin{tabular}[c]{@{}l@{}}top: 82\\ base: 84 \end{tabular}     \\ \hline
  \end{tabular}
We compare the design and measured (by SEM) dimensions on each face of the cleaved wafer. The top and base dimensions of the posts and holes are the width at the top of the structures (toward vacuum) and at the base (toward bulk silicon). The letters (B, C, T1, T2) refer to the dimensions in Fig.~\ref{fig:2lay_schem}. All layers use a 125~$\mu$m cell size (grid spacing; dimension A).  The thickness of the wafer is 996~$\mu$m, which is within the range provided by the manufacturer, Waferpro\textsuperscript{\textregistered}, ($1000 \pm 10$~$\mu$m).
  \end{threeparttable}
  \label{tab:dims_2lay_vals}
\end{table}

Fig.~\ref{fig:data_2layer} shows the reflectance and transmittance results of one of the tested samples, compared with both the original and updated HFSS simulations.  The sample shows $<$$-20$~dB reflectance over the band 187--317~GHz, meeting our 190--310~GHz design goal.  The measured reflectance agrees very closely with the simulation incorporating the SEM measurements, with discrepancies only below $-20$~dB.  An alternate technique that measures reflectance at normal incidence, not described here, agrees with these measurements to better than 1~dB precision after correcting for the incidence angle (black crosses in Fig.~\ref{fig:data_2layer}).

The remaining discrepancies with the simulation probably arise from small nonidealities created by the etching process.  We modeled some of these effects in Section~\ref{sS:Tolerancing} to characterize their magnitude, but we expect our model for these effects is only approximate.  In addition, because the HFSS simulations assumed periodic boundary conditions, they could not account for systematic or random variations with position.  In particular, we determined that the etch depth varies by $\pm$10\% from the nominal value, shallower at the center and deeper at the edge, and the taper angle of the walls also varies slightly with radius, from 0\textdegree\ at the center to no more than 1\textdegree\ at the edge, though we have not precisely characterized the range.  All these unmodeled nonidealities seem to only produce features below our $-20$~dB specification and thus do not merit further modeling.

The measurements do show a roughly 8~GHz shift relative to the {\em original} HFSS model, as measured below 100~GHz (the shift must be measured well away from the desired passband because the phasing of the Fabry-P\'erot fringing can significantly move the passband edges, as defined by the $-20$~dB transmission points).  This shift is largely due to the differences between the etched and design dimensions ($\pm 8\%$).  The second tested wafer showed the same shift, so we believe the differences are systematic and reproducible.  We are therefore confident we can reduce this shift substantially by accounting for these systematic differences with modest changes to the photolithographic masks and etch times.

The deviations from unity {\em transmittance} are likely due to measurement systematics, motivated by the observation that the deviations are not monotonically increasing with frequency.  We described in Section~\ref{S:test_setup} systematic effects in the test setup at this level.  The obvious alternative explanations, loss and scattering, would not yield such non-monotonic behavior.  Moreover, as we explained quantitatively in Section~\ref{S:Fabrication}, the loss expected given the high resistivity of the bulk silicon is much smaller than the observed deviation from unity transmittance.  It may be possible to use alternative measurement techniques to reduce these transmittance systematics, but they do not have a significant effect on the reflectance measurement.


\section{Impact of Wafer Bonding on Performance}
\label{S:waferbonding_results}

\begin{figure}[t!]
    \centering
    \includegraphics[width=\linewidth]{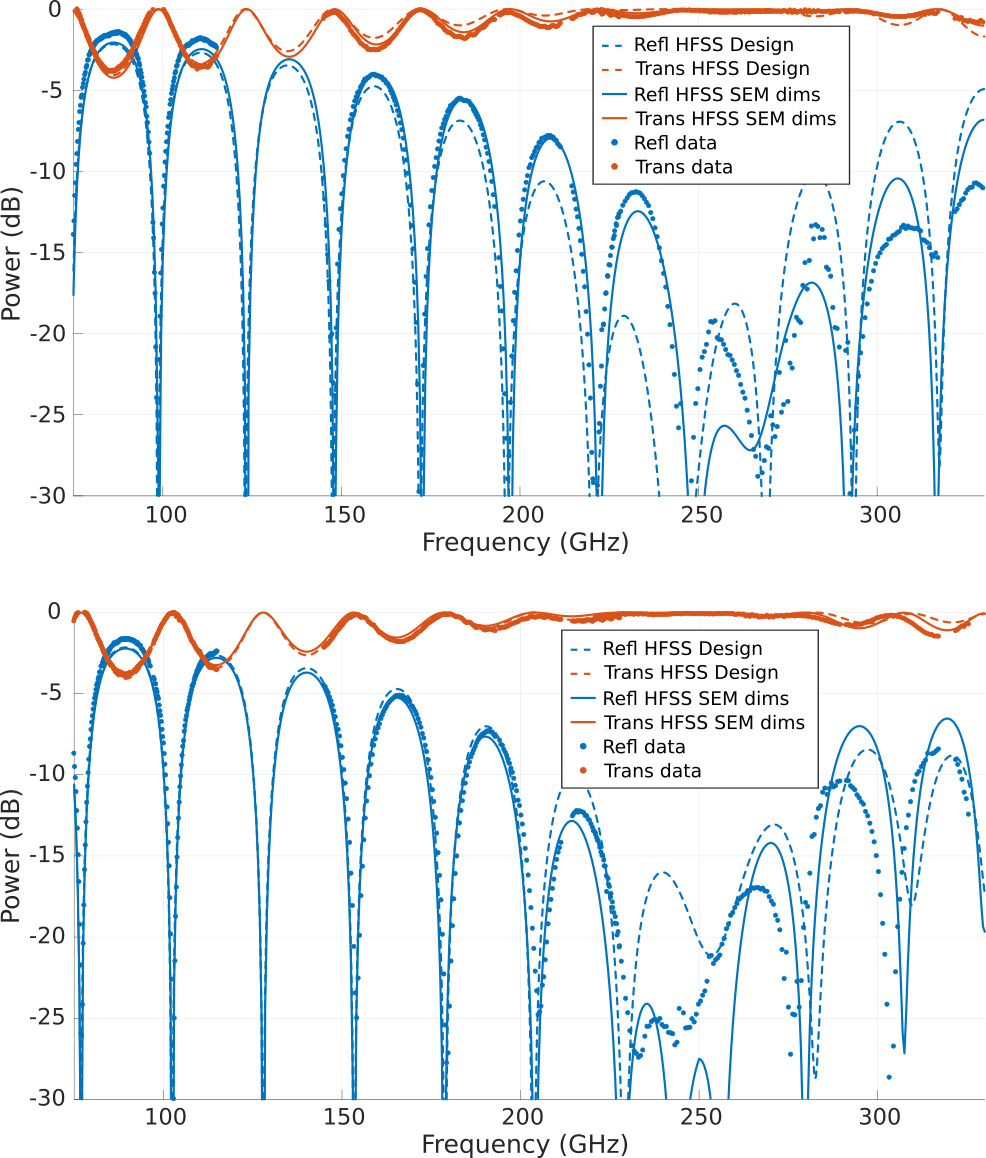}
    \caption{
    Reflectance and transmittance measurements of one-layer AR structures on two bonded silicon wafers compared to finite-element calculations using the design (dashed line) and measured dimensions (plain line) (Table~\ref{tab:dims1_lay_sem}), for a 15\textdegree angle of incidence.  (Top) Square posts. (Bottom) Square holes. }
    \label{fig:wafer-bonded_plots}
\end{figure}

Our overall approach (Section~\ref{sS:approach}) to producing broadband, antireflection-textured, gradient-index silicon optics involves wafer-bonding of patterned wafers using the technique described in Section~\ref{sS:waferbonding}.  As a first step toward demonstration of the process with etched wafers, we fabricated wafers with one-layer AR structures on one side (both the hole and post designs of Section~\ref{S:1lay_results}), bonded the {\it unpatterned} faces together, and tested their performance.  We show the bonded interface in Fig.~\ref{fig:waferbond} and the reflectance and transmittance results in Fig.~\ref{fig:wafer-bonded_plots}. 

The sample using holes shows excellent performance: the bandwidth over which we observe $<$$-$20~dB reflectance is comparable to that of the sample employing a one-layer structure on both sides of a single wafer (Fig.~\ref{fig:data_1layer}).  The agreement with theory is also good, though the Fabry-P\'erot pattern shows some deviations in the frequency band of interest.  The sample using posts does not perform as well, with a Fabry-P\'erot interference maximum rising up above $-$20~dB near the center of the desired band.  The measured dimensions partly explain the degraded performance.  Even then, the Fabry-P\'erot pattern shows the same kind of discrepancy in the desired band as we see in the wafer with holes.  These discrepancies are probably caused by variations of the structures' dimensions over the surface of the wafers, as explained in Section~\ref{S:2lay}.  We believe they are not germane to the quality of the wafer bond: it was observed in ~\cite{Gallardo:17} that a gap between wafers causes a distinctive, double-Fabry-P\'erot pattern, which we do not observe here.  We think it is reasonable to conclude that such effects, if present, are well below our $-$20~dB criterion.  Overall, this successful demonstration of wafer bonding is a necessary (but not sufficient) condition for bonding of patterned faces, our planned technique for structures requiring four or more layers, to yield acceptable performance.

\section{Application of Our Work to Future Development Efforts}
\label{S:next}


We have demonstrated the fabrication and optical performance of a two-layer, two-geometry (post and hole) antireflection structure on flat silicon suitable for THz applications.  While this technique provides a 1.6:1 bandwidth, which is sufficient to cover, for example, the 190--310~GHz atmospheric window, broader bandwidths and/or focusing optics are desirable for a range of applications. Our techniques can be expanded to achieve these goals by bonding together multiple wafers with such patterned structures. We have already demonstrated that adequate optical performance is maintained after wafer-bonding of unpatterned silicon surfaces.  Bonding of patterned surfaces is, therefore, a natural, although likely nontrivial, extension of our techniques. 

\begin{figure}[t!]
    \centering
    \includegraphics[width=\linewidth]{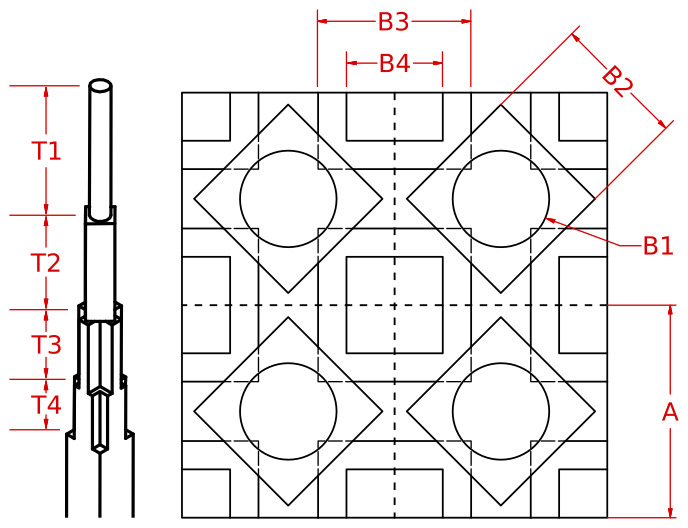}
	\caption{Schematic of a straw-person four-layer AR design, neglecting 
fabrication nonidealities. The features labeled T3 and T4 on the left would be patterned on one silicon wafer while the feature labeled T2 would be patterned on a separate silicon wafer. The two patterned surfaces would then be bonded and the feature labeled T1 would then be patterned to produce the final geometry. This AR structure would provide 4:1 bandwidth (see text).}
    \label{fig:4lay}
\end{figure}

To this end, we have designed a straw-person four-layer structure that would employ bonding of patterned wafers (Fig.~\ref{fig:4lay}).  HFSS calculations indicate it will have a 4:1 bandwidth, which would be sufficient to cover, for example, the atmospheric windows at 125--170~GHz, 190--310~GHz, and 335--355~GHz.  The top two layers use round and square posts, while the bottom two layers use square holes. The large etch depths of the post layers, along with the manner in which they intrude into the hole layers, make it impossible to fabricate the entire structure via multi-depth etching of a single wafer from the vacuum side.  Instead, we would pattern the two hole layers into a substrate silicon wafer via multi-depth etching as we have demonstrated here, pattern the higher-index post layer into a thin silicon wafer, bond the patterned faces together, and then etch the lower-index post layer into the vacuum side of the bonded structure. Even broader bandwidths should be possible using similar techniques. For example, a seven-layer design would be capable of covering a 6:1 bandwidth, which is sufficient for all of the atmospheric windows between 80 and 420~GHz . 

We anticipate the patterned structures we have demonstrated here can also be used to fabricate a gradient-index focusing optic. Obtaining adequate optical performance may require extreme aspect ratios for the features near the center and edges of the lens, and such features may be realizable by bonding patterned surfaces in the same manner as described above.  Based on our current straw-person designs, the bonding surfaces for some of the structures will likely be smaller than those used for the antireflection structures, and thus adequate bonding may be more challenging to realize.

\section{Conclusions}
\label{sec:conclusions}

We have successfully demonstrated a 1.6:1 bandwidth antireflection structure on a flat silicon wafer.  A silicon wafer patterned with this structure on both sides shows $<$$-$20~dB reflectance over the spectral band 187--317~GHz at 15\textdegree\ angle of incidence in TE polarization.  We believe observed deviations from unity transmission are not due to loss or scattering but rather to measurement systematics.  We have also demonstrated that wafer-bonding of unpatterned faces introduces no degradation in optical performance observable above the $-$20~dB level (also in TE at 15\textdegree).  These are important steps in the development of broadband, antireflection-textured, gradient-index optics.

\paragraph*{Funding.} 
NASA (NNX15AE01G)

\paragraph*{Acknowledgment.}  
We performed this work at the California Institute of Technology, the Caltech Submillimeter Observatory Hilo office, the Harvard-Smithsonian Center for Astrophysics, and the MicroDevices Laboratory of the Jet Propulsion Laboratory (operated by the California Institute of Technology under a contract with the National Aeronautics and Space Administration).  

The authors thank A.~Bose for early, pathfinding HFSS simulation work, K.~McClure for contributions to the HFSS tolerancing simulations, K.~Yee for performing the wafer-bonding steps, J.~Wong for contributions to the test setup control code, E.~Padilla for undertaking preparatory measurements of the two-layer structures, C.-Y.~E. Tong for participation in the Fig.~\ref{fig:data_2layer} alternate technique measurements, and T.~Macioce for contributions to the text of the paper.
C.~de~Young acknowledges support from an SAO Internship.  D.~Bisel, K.~Deniston, and S.~Stoll provided able administrative support.

\bibliography{bibliography}

\end{document}